\documentclass[journal]{IEEEtran}
\usepackage{cite}
\usepackage{amsmath,amssymb,amsfonts}
\usepackage{algorithmic}
\usepackage{graphicx}
\usepackage[bookmarks]{hyperref}
\usepackage{color}
\usepackage{textcomp}
\usepackage[utf8]{inputenc}
\usepackage{textgreek}
\usepackage{color}
\usepackage{cleveref}
\usepackage{url}
\usepackage{subfigure}
\usepackage{textcomp}
\usepackage{rotating}

\def\BibTeX{{\rm B\kern-.05em{\sc i\kern-.025em b}\kern-.08em
    T\kern-.1667em\lower.7ex\hbox{E}\kern-.125emX}}
\makeatletter
  \def\hrulefill{\leavevmode\leaders\hrule height 1pt\hfill\kern\z@}
\makeatother
\usepackage{wrapfig}
\usepackage{listings}

\definecolor{gris245}{RGB}{245,245,245}
\definecolor{olive}{RGB}{50,140,50}
\definecolor{brun}{RGB}{175,100,80}
\usepackage{fancybox}
\makeatletter
\newenvironment{CenteredBox}{%
\begin{Sbox}}{
\end{Sbox}\centerline{\parbox{\wd\@Sbox}{\TheSbox}}}
\makeatother
\def\BibTeX{{\rm B\kern-.05em{\sc i\kern-.025em b}\kern-.08em
    T\kern-.1667em\lower.7ex\hbox{E}\kern-.125emX}}
    
\def \post {{post-beamforming GLRT detector}}

\def \pre {{pre-beamforming GLRT detector}}
    
\begin{document}

\title{New Findings on GLRT Radar Detection of Nonfluctuating Targets via
Phased Arrays}

\author{Fernando Darío Almeida García, Marco Antonio Miguel Miranda and José Cândido Silveira Santos Filho \thanks{F.~D.~A.~García and J.~C.~S.~Santos Filho are with the Wireless Technology Laboratory, Department of Communications, School of Electrical and Computer Engineering, University of Campinas, 13083-852 Campinas, SP, Brazil, Tel.: +55 (19) 3788-5106, E-mails: $\{\text{ferdaral,candido}\}$@decom.fee.unicamp.br. } 
\thanks{M.~A.~M.~Miranda is with EMBRAER, Campinas, Brazil, Tel.: +55 19 2101-8800, E-mail: marco.miranda@embraer.com.br.
This work was supported by Coordenação de Aperfeiçoamento de Pessoal de Nível Superior (CAPES), Brazil, and by Secretaría de Educación Superior, Ciencia, Tecnología e Innovación (SENESCYT), Ecuador.
}
} 

\maketitle

\begin{abstract} 
This paper addresses the standard \textit{generalized likelihood ratio test} (GLRT) detection problem of weak signals in background noise. In so doing, we consider a \textit{nonfluctuating} target embedded in complex white Gaussian noise (CWGN), in which the amplitude of the target echo and the noise power are assumed to be unknown.
Important works have analyzed the performance for the referred scenario and proposed GLRT-based detectors. Such detectors are projected at an early stage (i.e., prior to the formation of a post-beamforming scalar waveform), thereby imposing high demands on hardware, processing, and data storage.
From a hardware perspective, most radar systems fail to meet these strong requirements. In fact, due to hardware and computational constraints, most radars use a combination of analog and digital beamformers (sums) before any estimation or further pre-processing.
The rationale behind this study is to derive a GLRT detector that meets the hardware and system requirements.
In this work, we design and analyze a more practical and easy-to-implement GLRT detector, which is projected after the analog beamforming. 
The performance of the proposed detector is analyzed and the probabilities of detection (PD) and false alarm (PFA) are derived in \textit{closed form}.
An alternative fast converging series for the PD is also derived.
This series proves to be very efficient and computationally tractable, saving both computation time and computational load.
Moreover, we show that in the low signal-to-noise ratio (SNR) regime, the \post{} performs better than both the classic \pre{} and the square-law detector.
This finding suggests that if the signals are weak, instead of processing the signals separately, we first must to reinforce the overall signal and then assembling the system's \textit{detection statistic}.
We also showed that the PFA of the \post{} is independent of the number of antennas. 
This property allows us to improve the PD (by increasing the number of antennas) while maintaining a fixed PFA.
At last, the SNR losses are quantified, in which the superiority of the \post{} was evidenced as the number of antennas and samples increase.
\end{abstract}


\maketitle

\section{Introduction}
Before performing any task (i.e., searching, tracking or imaging), the radar must decide whether the target of interest is present or absent in a certain range, angle or Doppler bin~\cite{blake86}. 
Unfortunately, the presence of unwanted signals such as thermal noise, clutter, and jamming, ubiquitous in practice,
often render this decision very complicated.
The optimal decision is achieved by applying the \textit{likelihood ratio test} (LRT)~\cite{leon94}. This decision is based on the Neyman-Pearson (NP) criterion, which maximizes the probability of detection (PD) for a given probability of false alarm (PFA)~\cite{chernoff54}. 
The LRT provides an optimal decision if the probability density functions (PDFs) of the received samples are fully known.
Of course, this requirement does not fit most practical problems.
In view of this, a more general decision rule arose to deal with these types of scenarios, the so-called \textit{generalized likelihood ratio test} (GLRT)~\cite{kay93}.
In the GLRT, all unknown PDF parameters are replaced by their maximum likelihood estimates (MLEs). 
This structure allows the GLRT to work over a wide range of scenarios.
Although, there is no optimality associated with the GLRT, in practice, it appears to work quite well.  

Important GLRT-based detectors were derived considering phased array radars, \textit{nonfluctuating} targets and, complex white Gaussian noise (CWGN) have been rigorously analyzed in the literature (cf.~\cite{kendall79,conte00,kay98,Almeida19,Haykin92} for more discussion on this).
These works assumed a partial or a complete lack of knowledge about the target and noise statistics. 
More complex detectors that rely on the use of secondary data can be found in~\cite{Haykin92,kelly86,Reed74,bose96,besson17,Pulsone01,Robey92}.
In these works, secondary data was assumed to be signal-free from the target components. That is, only noise is present.
In particular, in~\cite{kelly86}, it was derived the so-called Kelly’s detector, which considered that the primary and secondary data vectors share the same unknown noise covariance matrix.
In~\cite{besson17}, the authors extended the analysis by considering that the target amplitude follows a Gaussian distribution.

All referred works formulate the detection problem at an early stage (i.e., prior to the formation of a post-beamforming scalar waveform), thereby imposing high demands on hardware, processing and data storage.
In fact, due to hardware and computational constraints, most radars and mobile applications use a combination of analog and digital beamformers (sums) before any estimation or further pre-processing~\cite{Zhang18,Huber12,Axelsson03,Zhu17}. 
Furthermore, since the use of GLRT involves a high degree of mathematical complexity, theoretical performance analysis can be hampered in most situations.
Indeed, this was the case for the aforementioned studies in which their performance metrics -- probability of detection (PD) and probability of false alarm (PFA) -- were computed through numerical integration, estimated via Monte-Carlo simulations, expressed in integral-form, or require iterative solutions.
In this context, we also dedicate our efforts to easy the computation of the performance metrics.

Scanning the technical literature, we realize that no study has been devoted to the development of GLRT radar detectors using a post-beamforming approach.
In this paper, we design and evaluate a new GLRT-based detector which is projected after the analog beamforming operation.
Moreover, we provide the analytical tools to properly determine the performance of this detector. Specifically, we derive the PD and PFA in \textit{closed form}.
An alternative fast converging series for the PD is also derived. 
For the analysis, we consider a \textit{nonfluctuating} target embedded in CWGN, in which the amplitude of the target echo and the noise power are assumed to be unknown.
The use of secondary data is not considered.
From a mathematical point of view, one could envisage that our detector will somehow provide poorer performance since we are reducing the detection problem dimensionality by means of a sum operation (beamformer).
In this paper, we claim that this is not always the case if the signals are weak.
In fact, we show that in the low SNR regime, the \post{} performs better than the classic GLRT detector (called here as \pre{})~\cite[Eq. (6.20)]{kay98} and than the square-law detector~\cite[Eq. (15.57)]{richards10}, widely used in non-coherent radars~\cite{Weinberg17trans,Weinberg19,Weinberg17letter}.
This assertion suggest that, instead of processing the signals separately, it is better to adding them up before building the system's \textit{detection statistic}.
Other attractive features about our detector will be discussed throughout this work. 

The key contributions of this work may now be summarized as follows:
\begin{enumerate}
    \item Firstly, we design and evaluate a new GLRT detector projected after the analog beamforming operation. From the practical point of view, this detector meets the hardware and systems requirements of most radar systems.   
    \item Secondly, we obtain \textit{closed-form} expressions for the corresponding PD and PFA. In particular, the PD is given in terms of the bivariate Fox's $H$-function,  for which we also provide a portable and efficient MATHEMATICA routine.
    \item Thirdly, we derive an alternative series representation for the PD, obtained by exploring the orthogonal selection of poles in the Cauchy's residue theorem. This series enjoys a low computational burden and can be quickly executed in any ordinary desktop computer.\footnote{Section~\ref{sec:numerical_results} illustrates the efficiency of this series and compares it with MATHEMATICA's built-in numerical integration.}
    \item Finally, we provide some insightful and concluding remarks on the GLRT-based detection for \textit{nonfluctuating} targets.   
    To do so, we compare the performance of our derived detector with the pre-beamforming GLRT detector.
\end{enumerate}

The remainder of this paper is organized as follows.
Section~\ref{sec:Outline of the Radar Operation} describes the operation mode of our phased array radar. Section~\ref{sec:Phased Array Detection} describes the operation mode of the phased array radar. Section~\ref{sec:Performance Analysis} characterizes the \textit{detection statistics} and analyzes the corresponding performance metrics. Section~\ref{sec:Probability of Detection} introduces the multivariate Fox's $H$-function and derives both a \textit{closed-form} solution and a series representation for the PD.
Section~\ref{sec:numerical_results} discusses representative numerical results. Finally, Section~\ref{sec:conclusions} draws the main conclusions. 

In what follows, $f_{(\cdot)}(\cdot)$ denotes PDF; $\left( \cdot\right)^T$, transposition; $\left| \cdot \right|$, modulus; $\mathbf{Re} \left[ \cdot \right]$, real argument; $\mathbf{Im} \left[ \cdot \right]$, imaginary argument; $\left\| \cdot \right\|$, Euclidean norm; $\mathbb{E}\left[ \cdot \right]$, expectation; $\mathbb{COV}\left[ \cdot \right]$, covariance; $\text{rank} (\cdot) $, rank of a matrix; and $\left( \cdot\right) ^{-1}$, matrix inversion.

\section{Receiver's Front--End: Phased Array}
\label{sec:Outline of the Radar Operation}
\begin{figure}[t]
\begin{center}
\includegraphics[scale=0.40, clip, trim={8.65cm 0cm 9.8cm 0cm}, clip]{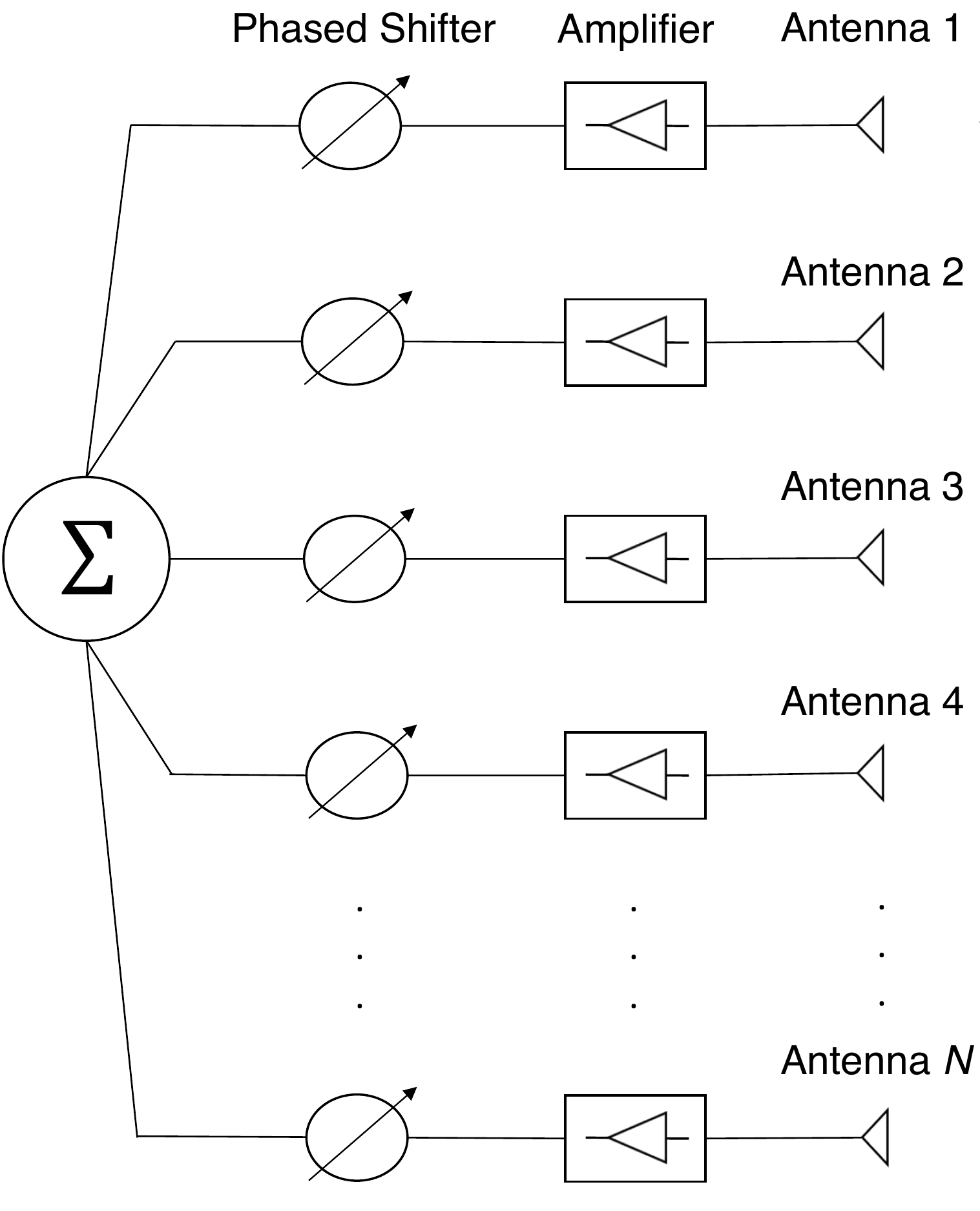}
\caption{Top view of the phased array radar.}
\label{fig:PhasedArrayScheme}
\end{center} 
\end{figure}

In this work, we consider a linear phased array radar composed of $N$ antennas equally separated in the azimuth direction, as shown in Fig.~\ref{fig:PhasedArrayScheme}. The transmission and reception processes are carried out as follows. A single antenna transmits a linear frequency-modulated pulse, whereas all antennas receive the echo signals.
Furthermore, an amplification block and a phased shifter are installed after each antenna element, and all outputs are added together (i.e., the analog beamforming operation is applied).

Thus, the in-phase and quadrature signals can be written in matrix form, respectively, as
\begin{align} 
\label{eq:x_vector}
\textbf{X}\triangleq & \left(
\begin{array}{cccc}
 X_{1,1} & X_{2,1} & \cdots  & X_{N,1} \\
 X_{1,2} & X_{2,2} & \cdots  & X_{N,2} \\
 \vdots  & \vdots  & \ddots & \vdots  \\
 X_{1,M} & X_{2,M} & \cdots  & X_{N,M} \\
\end{array}
\right)\\ \label{eq:y_vector}
\textbf{Y}\triangleq &\left(
\begin{array}{cccc}
 Y_{1,1} & \ Y_{2,1} & \cdots  & \ Y_{N,1} \\
 Y_{1,2} & \ Y_{2,2} & \cdots  & \ Y_{N,2} \\
 \vdots  & \ \vdots  & \ddots & \ \vdots  \\
 Y_{1,M} & \ Y_{2,M} & \cdots  & \ Y_{N,M} \\
\end{array}
\right), 
\end{align}
where $X_{n,m}$ and $Y_{n,m}$ represent the in-phase and quadrature received signals, respectively. In addition, $m \in \left\{1,2,\hdots,M \right\}$ is a discrete-time index, and $ n \in \left\{1,2,\hdots,N\right\}$ is a spacial index that denotes the association to the $n$-th antenna. 

For simplicity and without loss of generality, we assume a unity gain and a null phase shift for all antenna elements. In addition, we consider a collection of $M$ signal samples for each of the $N$ antennas. Then, the overall received signal can be written, in vector form,~as
\begin{align}
    \label{}
    \underline{R}= \left[R_1,R_2, \cdots,R_M \right]^T,
\end{align}
where
\begin{equation}
\label{eq:S}
R_m =\sum _{n=1}^N \left(X_{n,m}+j Y_{n,m}\right).
\end{equation}
Note that $\underline{R}$ is a complex-valued random vector, in which each component is formed by the sum of the received signals coming from  all the antennas at a certain time.

As will be shown in Section \ref{sec:Phased Array Detection}, the fact of adding the target echoes will drastically change the hardware design, \textit{detection statistic}, and performance of the \post{} compared to previous detectors (cf.~\cite{kay98,Haykin92,kelly86,bose96,besson17}). 
Since our detector is projected after the analog beamforming operation, one could argue that its performance would be somehow suboptimum, as compared to the \pre{}.
In this work, we show that this conclusion not always holds.
Indeed, for some cases the \post{} overcomes the \pre{}. This assertion heavily relies on the SNR of the incoming signals.

\section{Detection Design Via Post--Beamforming GLRT}
\label{sec:Phased Array Detection}
In this section, we present the detection scheme for the \post{}.

Herein, the presence of absence of the target is posed over the following \textit{binary hypothesis test}.\footnote{A \textit{binary hypothesis test} refers to the choice that a radar makes between two hypotheses: signal plus interference or only interference. This choice is made throughout all resolution cells~\cite{skolnik01}.}

\subsection{Hypothesis Test}
\label{sec:HypotesisTest}

\begin{itemize}
\item Hypothesis $\mathcal{H}_0$: target is absent. In this case, from the radar model described in the previous section, 
each $X_{n,m}$ and $Y_{n,m}$ are formed by mutually independent Gaussian components with zero mean and unknown variance $\sigma^2$. (Due to the presence of CWGN alone.) 

\item Hypothesis $\mathcal{H}_1$: target is present. In this case, 
each $X_{n,m}$ and $Y_{n,m}$ are formed by mutually independent Gaussian components with unknown non-zero means and unknown variance $\sigma^2$. (Due to the \textit{nonfluctuating} target and noise.) 

\end{itemize}
According to the stochastic model described in Section~\ref{sec:Outline of the Radar Operation}, the PDF of $\underline{R}$ under $\mathcal{H}_0$ is given by
\begin{align}
\label{eq:PDF_s_h0}
& \mathit{f}_{\underline{R}} \left(\underline{r}| \sigma^2;  \mathcal{H}_0 \right) =\frac{1}{\left(2 \pi \sigma^2 N \right)^M}  \exp \left[  -\frac{\sum _{m=1}^M \left| r_m \right|^2}{2  \sigma^2 N}\right],
\end{align}
whereas the PDF of $\underline{R}$ under $\mathcal{H}_1$ is given by~\eqref{eq:PDF_s_h1}, displayed at the top of the next page, where $\mu_{X}= \sum_{n=1}^{N}\mu_{X,n}$ and $\mu_{Y} =\sum_{n=1}^{N}\mu_{Y,n}$ represent the total sum of target echoes for the in-phase and quadrature components, respectively. 
Note that after the analog beamforming operation, we no longer have access to the specific value of target echo received by a particular antenna, which is what actually occurs in practice.
\begin{figure*}[ht]
\begin{flushleft}
\begin{align}
\label{eq:PDF_s_h1}
\mathit{f}_{ \underline{R}}  \left(\underline{r}| \sigma^2; \mu_X; \mu_Y; \mathcal{H}_1 \right)=\frac{1}{\left(2 \pi \sigma^2 N \right)^M}  \exp \left[  -\frac{\sum _{m=1}^M \left\{ \left(\mathbf{Re} \left[ r_m\right] - \mu_X \right)^2 +\left(\mathbf{Im} \left[ r_m\right] - \mu_Y \right)^2 \right\}}{2  \sigma^2 N}\right]
\end{align}
\normalsize
\end{flushleft}
\hrulefill
\end{figure*}

\subsection{Detection Rule}
The system's \textit{detection statistic} can be defined through GLRT as~\cite{kay98}
\begin{equation} \label{eq:GLRT}
\frac{f_{\underline{R}}\left(\underline{r}| \hat{\sigma}_1^2 ; \hat{\mu}_X; \hat{\mu}_Y;\mathcal{H}_{1} \right)}{f_{\underline{R}}\left(\underline{r}| \hat{\sigma}_0^2; \mathcal{H}_{0} \right)} \begin{array}{c} \mathcal{H}_1 \\ \gtrless \\ \mathcal{H}_0 \end{array} T,   
\end{equation}
\normalsize
where $T$ is an arbitrary threshold and the ratio on the left-hand side of~\eqref{eq:GLRT} is called the generalized likelihood ratio. In addition, $\hat{\sigma}_0^2$  is the MLE for $\sigma^2$, to be obtained from~\eqref{eq:PDF_s_h0}, and $\hat{\sigma}_1^2$, $\hat{\mu}_X$ and $\hat{\mu}_Y$ are the MLEs for $\sigma^2$, $\mu_X$ and $\mu_Y$, respectively, to be obtained from~\eqref{eq:PDF_s_h1}. Eq.\eqref{eq:GLRT} implies that the system will decide for $\mathcal{H}_1$ whenever the generalized likelihood ratio exceeds the threshold $T$, and will decide for $\mathcal{H}_0$ otherwise. 
Since the logarithmic function is a monotonically increasing function, we can rewrite the GLRT as
\begin{equation} \label{eq:GLRT_definition}
\ln  \left[  \frac{f_{\underline{R}}\left(\underline{r}| \hat{\sigma}_1^2 ; \hat{\mu}_X; \hat{\mu}_Y;\mathcal{H}_{1} \right)}{f_{\underline{R}}\left(\underline{r}| \hat{\sigma}_0^2; \mathcal{H}_{0} \right)} \right]  \begin{array}{c} \mathcal{H}_1 \\ \gtrless \\ \mathcal{H}_0 \end{array} \ln \left[ T\right].
\end{equation}
\normalsize
Note in~\eqref{eq:PDF_s_h0} and \eqref{eq:PDF_s_h1} that all unknown parameters $\left( \sigma^2, \mu_X \  \text{and} \  \mu_Y \right)$ are scalars quantities. Hence, the corresponding MLEs can be obtained easily. 
For example, $\hat{\sigma}_0^2$ can be found by taking the natural logarithm of~\eqref{eq:PDF_s_h0}, and then taking the derivative with respect to $\sigma^2$, i.e.,
\begin{align} \label{eq:log_sigma0}
\frac{\partial \ln \left[ \mathit{f}_{\underline{R}}  \left(\underline{r}| \sigma^2; \mathcal{H}_0  \right) \right]}{\partial \sigma^2} =  -\frac{M}{\sigma^2} +\frac{1}{2 N \sigma^4} \sum _{m=1}^M \left| r_m \right|^2.
\end{align}
\normalsize
Then, we set~\eqref{eq:log_sigma0} equal to zero and solve the equation for $\sigma^2$, which yields to
\begin{align} 
\label{eq:MLE_sigma0}
\hat{\sigma _0}^2  = &\frac{1}{2 M N} \sum _{m=1}^M \left| r_m \right|^2.
\end{align}
\normalsize
Using~\eqref{eq:PDF_s_h1} and following the same approach as in~\eqref{eq:MLE_sigma0}, the MLEs for $\mu_X$ and $\mu_Y$ can be calculated, respectively, as
\begin{align} \label{eq:MLE_AX}
\hat{\mu}_X =&\frac{1}{M }\sum _{m=1}^M \mathbf{Re} \left[ r_m\right]\\
\label{eq:MLE_AY}
\hat{\mu}_Y =& \frac{1}{M }\sum _{m=1}^M \mathbf{Im} \left[ r_m\right],
\end{align}
whereas the MLE for $\sigma^2$ can be computed as follows:
\begin{align}  \label{eq:MLE_sigma1}
\nonumber \hat{\sigma _1}^2 = & \frac{1}{2 N M} \sum _{m=1}^M \left\{ \left(\mathbf{Re} \left[ r_m\right] - \hat{\mu}_X\right)^2 \right.\\
& +  \left. \left(\mathbf{Im} \left[ r_m\right] - \hat{\mu}_Y\right)^2 \right\}.
\end{align}
(For brevity, we have omitted the derivation steps.)

Substituting~\eqref{eq:MLE_sigma0}--\eqref{eq:MLE_sigma1} in~\eqref{eq:GLRT_definition} and after simple simplifications, we have
\begin{align} \label{eq:lambdaS}
M \ln \left[ \left( \frac{\hat{\sigma _0}^2}{\hat{\sigma _1}^2}\right) \right] \begin{array}{c} \mathcal{H}_1 \\ \gtrless \\ \mathcal{H}_0 \end{array} \ln \left[ T\right].
\end{align}
\normalsize
Expanding~\eqref{eq:MLE_sigma1} and after performing some minor manipulations, we can rewrite $\hat{\sigma _1}^2$ as 
\begin{align}
\label{eq:sigma0_expand}
\nonumber \hat{\sigma _1}^2 & = \frac{1}{2 M N} \sum _{m=1}^M \left\{\hat{\mu}_X^2+\hat{\mu}_Y^2\right\} \\
\nonumber & + \underbrace{\frac{1}{2 M N} \sum _{m=1}^M \left\{ \left(\mathbf{Re} \left[ r_m\right] \right)^2 + \left(\mathbf{Im} \left[ r_m\right] \right)^2\right\}}_{\hat{\sigma _0}^2} \\
\nonumber & +   \left(\frac{\hat{\mu}_X}{N} \right) \underbrace{\frac{1}{M} \sum _{m=1}^M \mathbf{Re} \left[ r_m\right] }_{\hat{\mu}_X}+ \left(\frac{\hat{\mu}_Y}{N} \right) \underbrace{\frac{1}{M} \sum _{m=1}^M \mathbf{Im} \left[ r_m\right]}_{\hat{\mu}_Y}\\
 & \overset{(a)}{=} \hat{\sigma _0}^2 -\frac{1}{2 N} \left(\hat{\mu}_X^{2}+\hat{\mu}_Y^2\right),
\end{align}
where in step (a) we have used~\eqref{eq:MLE_sigma0}, \eqref{eq:MLE_AX}, and \eqref{eq:MLE_AY}, along with some simplifications. 

Isolating $\hat{\sigma}_0^2$ from ~\eqref{eq:sigma0_expand}, we obtain
\begin{align}
\label{eq:sigma0expanded}
\hat{\sigma}_0^2=\hat{\sigma}_1^2+\frac{1}{2 N} \left(\hat{\mu}_X^{2}+\hat{\mu}_Y^2\right).
\end{align}
\normalsize
Replacing~\eqref{eq:sigma0expanded} in~\eqref{eq:lambdaS}, yields
\begin{align} \label{eq:lambdaSv2}
M \ln \left[ 1+   \frac{ \left(\hat{\mu}_X^2+\hat{\mu}_Y^2\right)}{2 N \hat{\sigma _1}{}^2} \right] \begin{array}{c} \mathcal{H}_1 \\ \gtrless \\ \mathcal{H}_0 \end{array} \ln \left[ T\right].
\end{align}
\normalsize
Now, since $M$ and $N$ are a positive numbers, we obtain the same decision as in~\eqref{eq:lambdaSv2} by simply comparing $ \left( \hat{\mu}_X^2+\hat{\mu}_Y^2 \right) /  \hat{\sigma}_1^2$ with a modified threshold, $\gamma'$, that is,
\begin{equation} \label{eq:Psi}
 \frac{\hat{\mu}_X^2+\hat{\mu}_Y^2}{\hat{\sigma _1}^2} \begin{array}{c} \mathcal{H}_1 \\ \gtrless \\ \mathcal{H}_0 \end{array}  \gamma'.
\end{equation}
\normalsize
For convenience and without loss of generality, we define an equivalent decision rule as\footnote{The constant $\Psi$ was introduced in the decision rule because it allow us to model $Z$ as a random variable with known PDF, as will become apparent soon.}
\begin{align}
    \label{eq:Zfinal}
      Z \triangleq \Psi \left(  \frac{\hat{\mu}_X^2+\hat{\mu}_Y^2}{\hat{\sigma _1}^2}\right)  \begin{array}{c} \mathcal{H}_1 \\ \gtrless \\ \mathcal{H}_0 \end{array} \gamma,
\end{align}
\normalsize
where $Z$ is the system's \textit{detection statistic}, $\Psi = (M-1)/ 2 N$ is a positive constant, and $\gamma$ is a new modified threshold.

\begin{figure*}
  \centering
    \begin{tabular}[b]{c}
    \includegraphics[trim={0cm 0cm 0cm 0cm}, clip,scale=0.40]{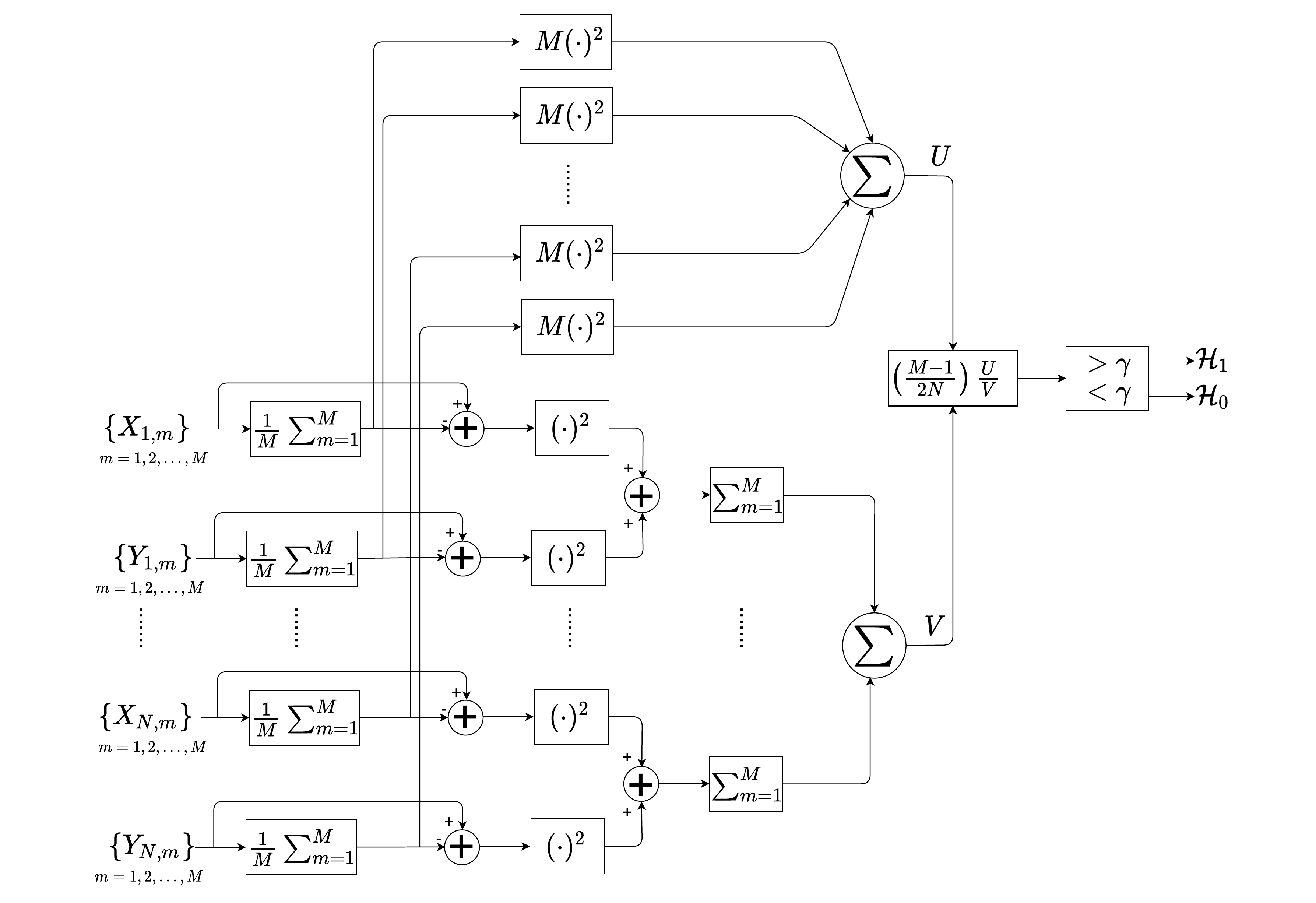} \\
   \textbf{(a)} Pre-beamforming GLRT detector~\cite{kay98}.
  \end{tabular} \qquad 
    \begin{tabular}[b]{c}
    \includegraphics[trim={1.5cm 0cm 2cm -1.5cm}, clip,scale=0.40]{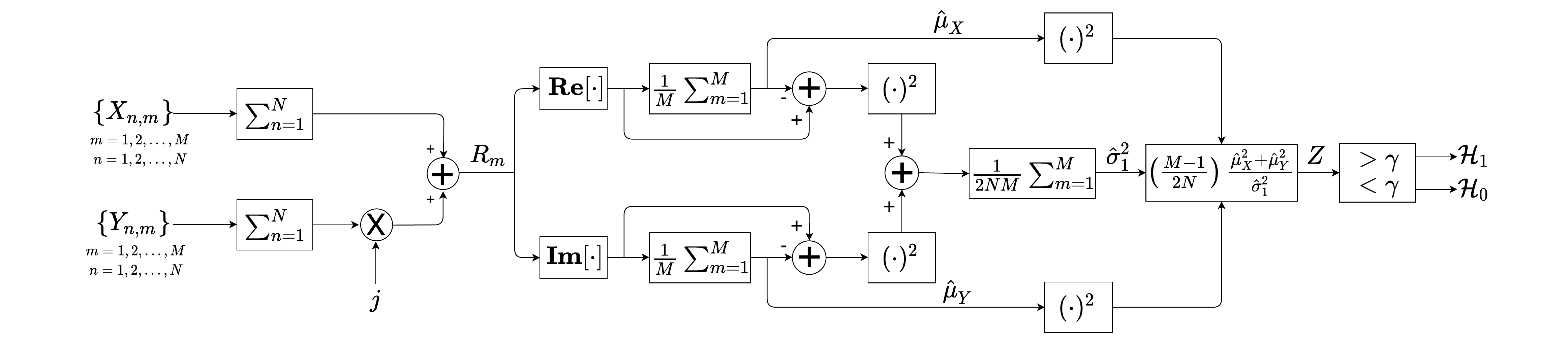}  \\
     \textbf{(b)} Post-beamforming GLRT detector.
  \end{tabular} \qquad 
    \begin{tabular}[b]{c}
    \includegraphics[trim={0cm 0cm 2cm -1.5cm}, clip,scale=0.40]{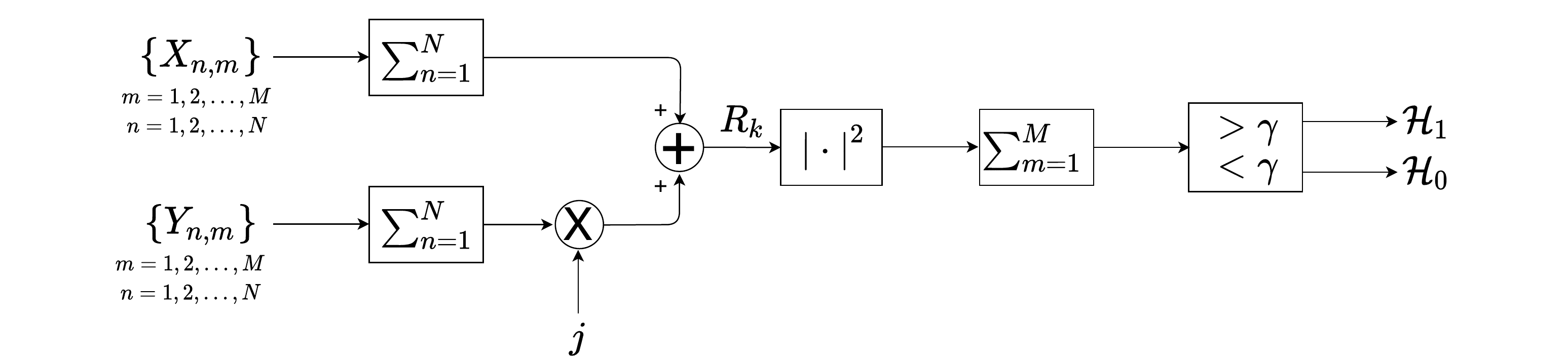}  \\
     \textbf{(c)} Square-law detector~\cite{richards10}.
  \end{tabular} 
  \caption{Detection Schemes.}
  \label{fig: Detector Schemes}
  \vspace{0cm}
  \hrulefill
\end{figure*}

Fig.~\ref{fig: Detector Schemes} illustrates how the pre-beamforming GLRT, the post-beamforming GLRT, and the square-law detectors are constructed.
More specifically, Fig.~\ref{fig: Detector Schemes}-(a) depicts the pre-beamforming GLRT detector architecture. In this case, all received signals are processed separately to form the system's \textit{detection statistic} \cite{kay98}. Certainly, this type of processing is more difficult to implement due to hardware constraints.
Fig.~\ref{fig: Detector Schemes}-(b) illustrates the \post{} architecture. This detector provides a less restrictive hardware implementation, as well as a simpler \textit{detection statistic} that results from adding the received signals. 
Finally, Fig.~\ref{fig: Detector Schemes}-(c) illustrates the square-law detector architecture. Here, after the analog beamforming, the square magnitude of the signal samples is taken and then they are added up together.
It is important to emphasize that in order to analytically calculate the performance metrics of the square law detector, we do need the information about the noise power. That is, for a given PFA, the detection threshold is given as a function of the noise power~\cite{richards10}. 

\section{Detection Performance}
\label{sec:Performance Analysis}
In this section, we characterize and analyze the performance of the \post{}. To do so, we start finding the PDFs of Z under $\mathcal{H}_0$ and $\mathcal{H}_1$.


\subsection{Detection Statistics}
\label{sec:Decision}
First, we rewrite~\eqref{eq:Zfinal} as follows: 
\begin{align}
    \label{eq:Z_Final}
    \nonumber Z&=  \frac{ (M-1) \left(\hat{\mu}_X^2+\hat{\mu}_Y^2\right)}{2 N \hat{\sigma _1}^2}\\
    & \overset{(a)}{=} (M-1) \frac{ \overbrace{ \left(\hat{\mu}_X^2+\hat{\mu}_Y^2\right) M / N \sigma^2 }^{\triangleq \ 
    \mathcal{I}_1} }{\underbrace{2 \hat{\sigma _1}^2 M / \sigma^2}_{\triangleq  \ \mathcal{I}_2} },
\end{align}
\normalsize
where in step (a), without affecting the detection performance, we have multiplied the left-hand side of $Z$ by $M \sigma^2/M\sigma^2$.

Note that, to fully characterize $Z$, it is imperative to find the PDFs of $\mathcal{I}_1$ and  $\mathcal{I}_2$ under $\mathcal{H}_0$ and~$\mathcal{H}_1$.

Substituting~\eqref{eq:MLE_AX} and \eqref{eq:MLE_AY} in $\mathcal{I}_1$, yields to
\begin{align}
    \label{eq:I1_final}
    \nonumber \mathcal{I}_1 = & \underbrace{\left(\frac{1}{\sqrt{M N} \sigma}\sum _{k=1}^M \mathbf{Re} \left[ r_k\right]  \right)^2}_{\triangleq \ U} \\
     &+  \underbrace{\left( \frac{1}{\sqrt{M N} \sigma}\sum _{k=1}^M \mathbf{Im} \left[ r_k\right]\right)^2}_{\triangleq \ V}.
\end{align}
\normalsize
Hereinafter, the detector in~\cite[Eq. (6.20)]{kay98} will be called Fox's $H$-function GLRT phased array detector.
Observe that $U$ is the square of a Gaussian random variable (RV) with mean $\sqrt{M} \mathbb{E}\left[ X_{l,k} \right] / \sigma \sqrt{N}$ and unit variance. In a similar way, $V$ is the square of a Gaussian RV with mean $\sqrt{M} \mathbb{E}\left[ Y_{l,k} \right] / \sigma \sqrt{N}$ and unit variance.
Therefore, depending on the hypothesis, $\mathcal{I}_1$  can match one of the following conditions:
\begin{enumerate}
    \item Given $\mathcal{H}_0$: $\mathcal{I}_1$ follows a central chi-squared (CCS) distribution~\cite{papoulis02} with $\nu_1 = 2$ degrees of freedom.
    \item Given $\mathcal{H}_1$: $\mathcal{I}_1$ follows a  noncentral chi-squared (NCCS) distribution~\cite{patnaik49} with noncentral parameter $\lambda_1 = M  \left(\mu_{X}^2+\mu_{Y}^2\right) / N \sigma^2$ and $\alpha_1 = 2$ degrees of freedom.
\end{enumerate}
Inserting~\eqref{eq:MLE_sigma1} in $\mathcal{I}_2$, we obtain
\begin{align}
\label{eq:I2_final}
\nonumber\mathcal{I}_2 =    \frac{1}{N \sigma^2} & \sum _{m=1}^M  \left\{\left(\mathbf{Re} \left[ r_m\right]- \hat{\mu}_X\right)^2 \right.\\
& + \left. \left(\mathbf{Im} \left[ r_m\right]- \hat{\mu}_Y\right)^2 \right\}
\end{align}
\normalsize
Here, the analysis is a bit more cumbersome; therefore, we establish the following two lemmas:

\medskip

\textit{Lemma 1}: $\mathcal{I}_2$ matches the following conditions:
\begin{enumerate}
    \item Given $\mathcal{H}_0$:  $\mathcal{I}_2$  follows a CCS distribution with $\nu_2 = 2(M-1)$ degrees of freedom.
    \item Given $\mathcal{H}_1$: $\mathcal{I}_2$ also follows a CCS distribution with  $2(M-1)$ degrees of freedom. In this case, for convenience, we model $\mathcal{I}_2$ by a NCCS distribution with noncentral parameter $\lambda_2 =0 $ and $\alpha_2 = 2(M-1)$ degrees of freedom.
\end{enumerate} 

\textit{Proof}: See Appendix~A.
\hspace*{\fill} $\blacksquare$ 
\medskip

\textit{Lemma 2}: $\mathcal{I}_1$ and $\mathcal{I}_2$ are mutually independent RVs. 
\medskip

\textit{Proof}: See Appendix~B.
\hspace*{\fill} $\blacksquare$ 

\medskip

Then, using Lemmas 1 and 2, we can define $\mathcal{I}_1/ \mathcal{I}_2$ as the ratio of either two independent CCS RVs or two independent NCCS RVs, depending on the hypothesis. 
The factor $(M-1)$ in~\eqref{eq:Z_Final} allows us to model $Z$ by a RV with known PDF.

Given $\mathcal{H}_0$, it can be shown that $Z$ follows a central F-distribution~\cite{phillips82} with  PDF given by
\begin{align}
\label{eq:PDF_h0_beta}
\mathit{f}_Z\left(z| \mathcal{H}_0\right)&=\frac{(M-1)^{M-1} (M+z-1)^{-M}}{B(1,M-1)},
\end{align}
\normalsize
where $B(\cdot,\cdot)$ is the Beta function~\cite[Eq. (5.12.3)]{Olver10}. 
Using~\cite[Eq. (5.12.1)]{Olver10}, we can rewrite~\eqref{eq:PDF_h0_beta} in compact form as
\small
\begin{align}
\label{eq:PDF_h0}
\mathit{f}_Z\left(z| \mathcal{H}_0\right)=\left(\frac{M-1}{M+z-1}\right)^M.
\end{align}
\normalsize

For the case of $\mathcal{H}_1$, $Z$ can be modeled by a doubly noncentral F-distribution~\cite{Bulgren71}, with PDF given by
\begin{align}
\label{eq:pdfH1_4}
\nonumber \mathit{f}_Z\left(z| \mathcal{H}_1\right)= & \exp \left[-\Upsilon \ M  \right] \left(\frac{M-1}{M+z-1}\right)^M   \\
& \times \, _1F_1\left(M;1;\frac{ \Upsilon \ z \  M  }{ M+z-1}\right),
\end{align}
\normalsize
where $\Upsilon=(\mu_X^2 +\mu_Y^2)/2 N \sigma^2 $, and $_1F_1\left(\cdot;\cdot;\cdot\right)$ is the Kummer confluent hypergeometric
function~\cite[Eq. (13.1.2)]{Olver10}.
The equality $\Upsilon= N \  \text{SNR}_n$ holds if $\text{SNR}_n=\text{SNR}_p \  \forall \ (n,p)$, with $\text{SNR}_n=\left( \mu_{X,n}^2+\mu_{Y,n}^2 \right) /2 \sigma^2$ being the signal-to-noise ratio present at the $n$-th antenna.
The derivation of~\eqref{eq:pdfH1_4} is shown in Appendix~C.

\subsection{False Alarm and Detection Probabilities}
\label{sec:Performance Metrics}

It is well known that the performance of any radar system is governed by the PFA and PD. These probabilities can be computed, respectively, as~\cite{skolnik01}
\begin{align}
\label{eq:def_pfa}
P_{\text{FA}} & \triangleq \int_{\gamma }^{\infty } \mathit{f}_Z\left(z| \mathcal{H}_0\right) \, \text{d} z \\ 
\label{eq:def_pd}
P_{\text{D}} & \triangleq \int_{\gamma}^{\infty } \mathit{f}_Z \left(  z| \mathcal{H}_1 \right) \, \text{d} z.
\end{align}
\normalsize
Replacing~\eqref{eq:PDF_h0} in~\eqref{eq:def_pfa}, yields
\begin{align}
\label{eq:PFA}
P_{\text{FA}} = \left(\frac{M-1}{\gamma +M-1}\right)^{M-1}.
\end{align}
\normalsize
Now, isolating $\gamma$ from~\eqref{eq:PFA} we can find a threshold so as to meet a desired PFA, i.e.,
\begin{align}
    \label{eq:gamma}
    \gamma =1-M+ \left( M-1\right) 
    {P_{\text{FA}}}^{1/(1-M)}.
\end{align}
\normalsize
It can be noticed in \eqref{eq:gamma} that we do not need the knowledge of the noise power nor the number of antennas to set the detection threshold. That is, the detection threshold $\gamma $ is independent of both $\sigma^2$ and $N$. This important feature will allow us to maintain a certain PFA for an arbitrary number of antennas. More precisely, with objective of increasing the PD, we can increase $N$ without worrying about the increase in the PFA.

On the other hand, after substituting~\eqref{eq:pdfH1_4} in~\eqref{eq:def_pd}, the PD can be obtained in single-integral form as
\begin{align}
    \label{eq:PD_Integral_terms}
    \nonumber P_{\text{D}} = & \exp \left[-\Upsilon \ M   \right] \int_{\gamma }^{\infty }  \left(\frac{M-1}{M+z-1}\right)^M   \\
    & \times \, _1F_1\left(M;1;\frac{ \Upsilon \ z \  M }{ M+z-1}\right)  \, \text{d} z.
\end{align}
\normalsize
Certainly, \eqref{eq:PD_Integral_terms} can be evaluated by means of numerical integration. 
Nonetheless, to further facilitate the computation of the PD, we provide alternative, faster, and more tractable solutions. 
This is attained in the next section.

\section{Alternative Expressions for the Probability of Detection}
\label{sec:Probability of Detection}
In this section, we provide both a closed-form solution and a fast converging series for the PD, 
To this end, we make use complex analysis and a thorough calculus of residues.

\subsection{The Multivariate Fox's $H$-function}
\label{subsec:The Multivariate Fox's $H$-function}
We first begin introducing the Fox's $H$-function, as it will be used throughout this section.

The Fox's $H$-function has been used in a wide variety of recent applications, including mobile communications and radar systems  (cf.~\cite{Garcia19,Rahama18,carlos17,hfox,AlmeidaIET20} for more discussion on this). In~\cite{hay95}, the authors considered the most general case of the Fox's $H$-function for several variables, defined as 
\begin{equation}
\label{eq:FoxDefinition1}
\mathbf{H} \left[ \textbf{x};\left(\delta, \textbf{D} \right);\left(\beta , \textbf{B} \right); \mathcal{L}_\textbf{s} \right] \triangleq \left(\frac{1}{2 \pi j} \right)^L \oint_{\mathcal{L}_{\textbf{s}}} \Theta \left(  \textbf{s} \right)\textbf{x}^{-\textbf{s}} \text{d} \textbf{s},
\end{equation}
\normalsize
in which $j=\sqrt{-1}$ is the imaginary unit, $\textbf{s}\triangleq\left[ s_1, \cdots,s_L  \right]$, $ \textbf{x}\triangleq \left[ x_1,\cdots,x_L   \right]$, $\beta\triangleq \left[ \beta_1,\cdots, \beta_L \right]$,  and $\delta \triangleq \left[  \delta_1, \cdots, \delta_L \right]$ denote vectors of complex numbers, and $\textbf{B}\triangleq \left(b_{i,j}  \right)_{n \times L}$ and $\textbf{D}\triangleq\left( d_{i,j} \right)_{m \times L}$ are matrices of real numbers. Also, $\textbf{x}^{-\textbf{s}}\triangleq \prod_{i=1}^{L} x_i^{- s_i}$, $\text{d} \textbf{s}\triangleq\prod_{i=1}^{L} \text{d} s_i$, $\mathcal{L}_{\textbf{s}}\triangleq\mathcal{L}_{\textbf{s},1} \times \cdots \times \mathcal{L}_{\textbf{s},L}$, $\mathcal{L}_{\textbf{s},k}$ is an  appropriate contour on the complex plane $s_k$, and
\begin{equation}
\label{eq:FoxDefinition2}
 \Theta \left(  \textbf{s} \right) \triangleq \frac{\prod _{i=1}^m \Gamma \left(\delta _i+\sum_{k=1}^L d_{i,k} s_k\right)}{\prod _{i=1}^n \Gamma \left(\beta _i+\sum_{k=1}^L b_{i,k} s_k\right)},
\end{equation}
\normalsize
in which $\Gamma (\cdot)$ is the gamma function~\cite[Eq. (6.1.1)]{abramowitz72}.
 

\subsection{Fox's H-Function-Based Representation}
Here, we obtain an alternative closed-form solution for \eqref{eq:PD_Integral_terms}, expressed in terms of the Fox's $H$-function.

To do so, we first perform some mathematical manipulations in \eqref{eq:PD_Integral_terms}, resulting in
\begin{align}
    \label{eq:PD Meijer}
    \nonumber P_{\text{D}} =& \frac{\exp \left[-  \Upsilon \  M\right] (M-1)^M}{\Gamma(M)} \int _{\gamma }^{\infty } \left(\frac{1}{M+z-1}\right)^M \\
    &\times  G_{1,2}^{1,1}\left[
 \left. \begin{array}{c}
 1-M \\
 0,0 \\
\end{array} 
 \right|-\frac{ \Upsilon \ z \ M}{M+z-1} 
\right]\text{d}z,
\end{align}
\normalsize
where $G_{m,n}^{p,q} \left[ \cdot \right]$ is the Meijer's G-function~\cite[Eq. (8.2.1.1)]{prudnikov03}.

Now, using the contour integral representation of the Meijer's G-function, we can express~\eqref{eq:PD Meijer} as follows:
\begin{align}
    \label{eq:}
    \nonumber  P_{\text{D}}  =& \frac{\exp \left[- \Upsilon  \ M \right] (M-1)^M}{\Gamma(M)}  \int _{\gamma }^{\infty }  \left(\frac{1}{M+z-1}\right)^M  \\
    \nonumber & \times \left( \frac{1}{2 \pi j}\right) \oint_{\mathcal{L}^{**}_{\textbf{s},1}}   \frac{\Gamma (s_1) \Gamma (M-s_1)}{\Gamma (1-s_1)} \\
    & \times \left(-\frac{ \Upsilon \ z \ M}{M+z-1}  \right)^{-s_1} \text{d} s_1 \ \text{d}z,
\end{align}
\normalsize
in which $\mathcal{L}^{**}_{\textbf{s},1}$ is a closed complex contour that separates the poles of the gamma function $\Gamma(s_1)$ from the poles of $\Gamma(M-s_1)$.
Since $\int _{\gamma }^{\infty } \left| \mathit{f}_Z  \left(z| \mathcal{H}_1\right) \right| \text{d}z < \infty$, we can interchange the order of integration\cite{fubibi07}, i.e.,
\begin{align}
    \label{eq:}
    \nonumber P_{\text{D}} =& \frac{\exp \left[-  \Upsilon \ M \right] (M-1)^M}{\Gamma(M)} \left( \frac{1}{2 \pi j}\right)  \\
    \nonumber & \times  \oint_{\mathcal{L}^{**}_{\textbf{s},1}}  \frac{\Gamma (s_1) \Gamma (M-s_1) \left(-  \Upsilon \ M \right)^{-s_1}}{\Gamma (1-s_1)}  \\
    & \times \int _{\gamma }^{\infty } \left(\frac{1}{M+z-1}\right)^{M} \left(\frac{z}{M+z-1}  \right)^{-s_1}  \text{d}z \ \text{d} s_1.
\end{align}
\normalsize
Developing the inner real integral, we obtain
\begin{align}
    \label{eq:PD regularized}
    \nonumber P_{\text{D}} =& \frac{\exp \left[- \Upsilon \ M \right] (M-1)^{M}   \Gamma (M-1) }{ \Gamma (M) \  \gamma^{M-1} } \left( \frac{1}{2 \pi j}\right) \\
    \nonumber & \times   \oint_{\mathcal{L}^{*}_{\textbf{s},1}}  \frac{\Gamma (s_1) \Gamma (M-s_1) \left(-  \Upsilon \ M \right)^{-s_1}}{\Gamma (1-s_1)} \\
    & \times \, _2\tilde{F}_1\left(M-1,M-s_1;M;\frac{1-M}{\gamma }\right) \text{d} s_1,
\end{align}
\normalsize
where $\, _2\tilde{F}_1(a,b;c;x)=\, _2F_1(a,b;c;x)/ \Gamma (c)$ is the regularized Gauss hypergeometric function, and $\, _2F_1(\cdot,\cdot; \cdot;\cdot)$ is the Gauss hypergeometric function~\cite[Eq. (15.1.1)]{Olver10}. 
Note that we have used a new complex contour, $\mathcal{L}^{*}_{\textbf{s},1}$. This is because the inner integration changed the integration path in the complex plane. 
Here, $\mathcal{L}^{*}_{\textbf{s},1}$ is a closed contour that separates the poles of $\Gamma(s_1)$  from those of $\Gamma(M -s_1)$.

\begin{figure}[t]
\begin{center}
\includegraphics[scale=0.43, clip, trim={0cm 0cm 5cm 0cm}, clip]{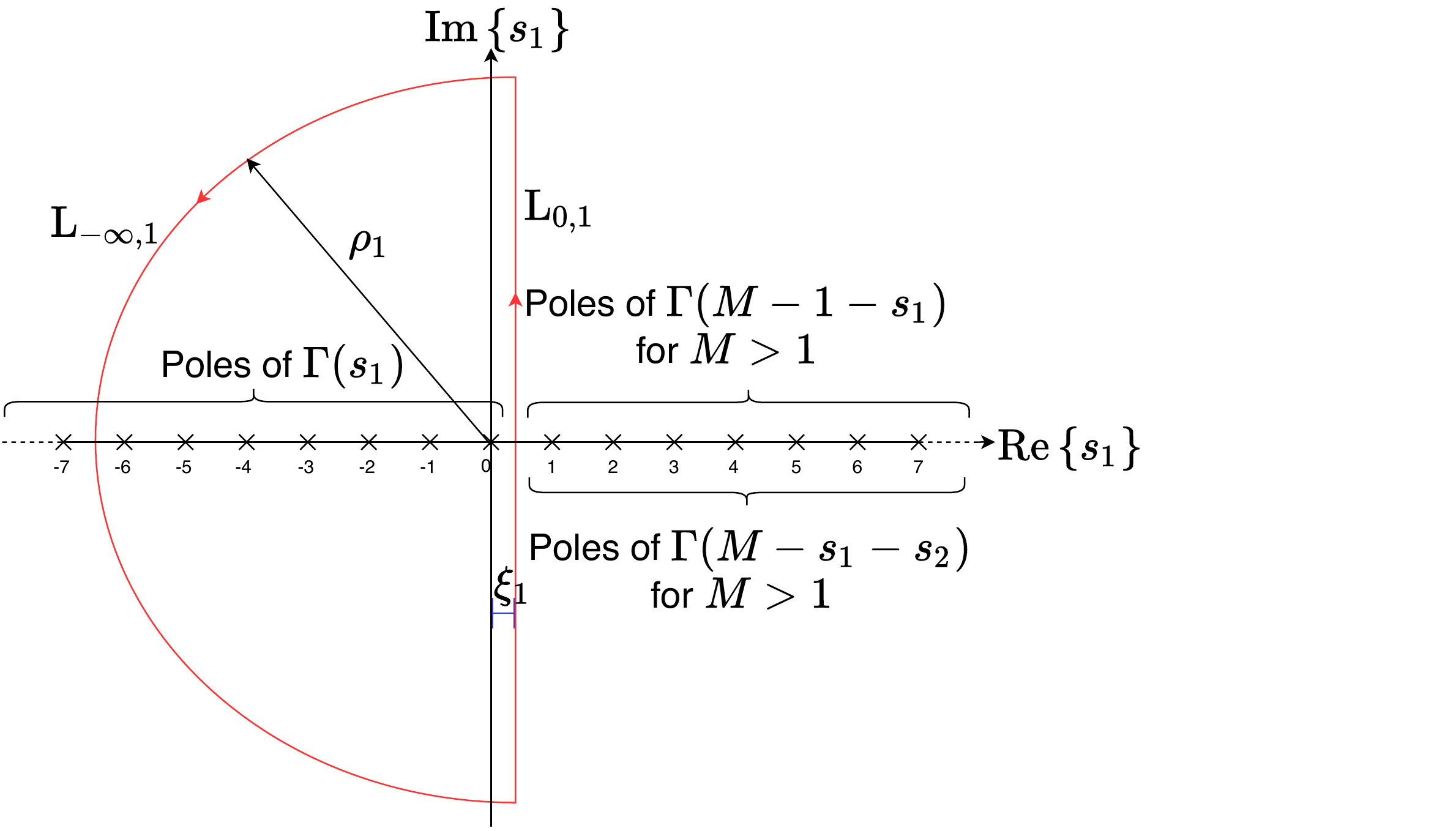}
\caption{Integration path for $\mathcal{L}_{\textbf{s},1}$.}
\label{fig: L1}
\end{center} 
\end{figure}
\begin{figure}[t]
\begin{center}
\includegraphics[scale=0.43, clip, trim={0cm 0cm 5cm 0cm}, clip]{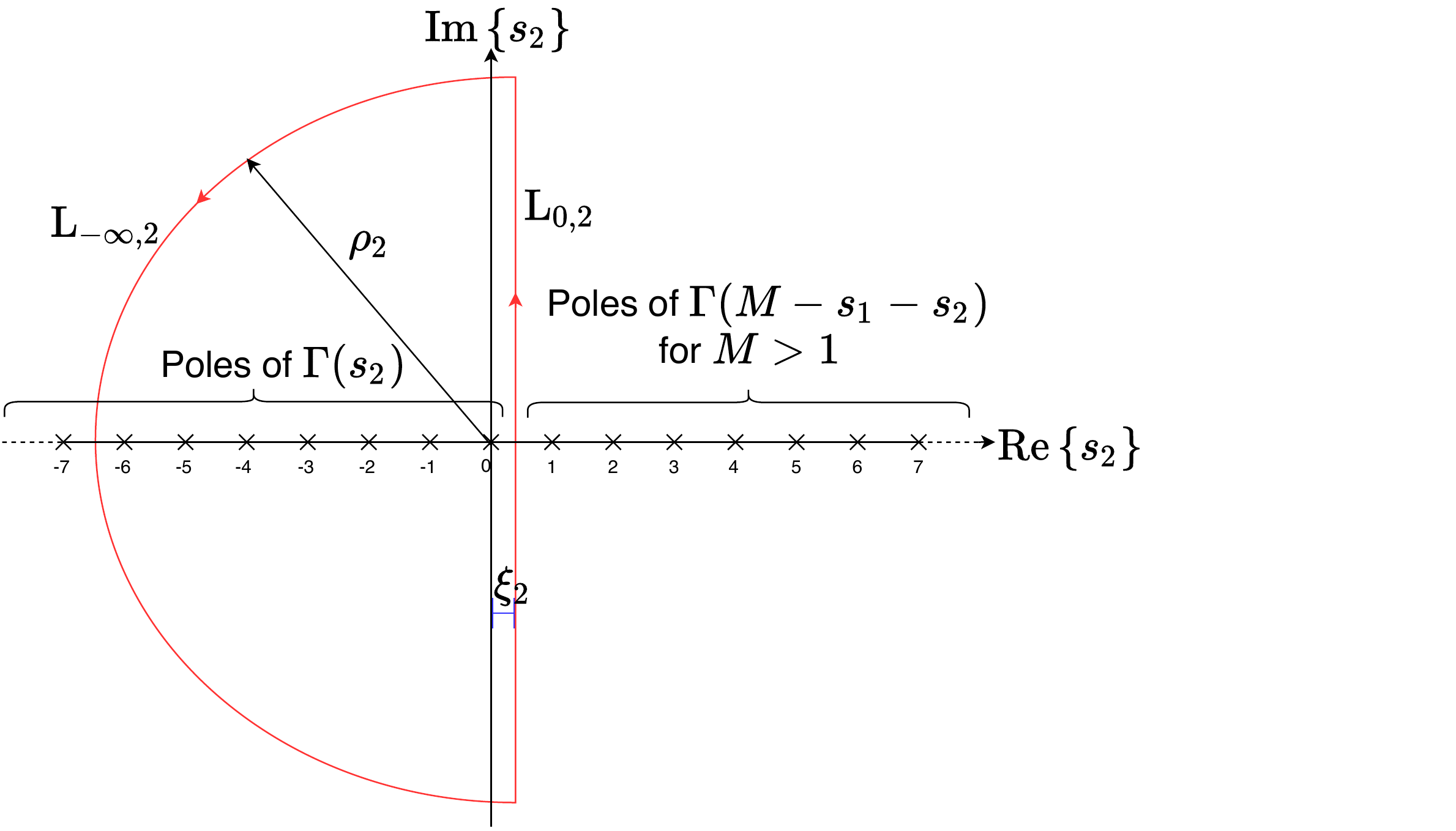}
\caption{Integration path for $\mathcal{L}_{\textbf{s},2}$.}
\label{fig: L2}
\end{center} 
\end{figure}
Finally, replacing \eqref{eq:gamma} in~\eqref{eq:PD regularized} and after using the complex integral representation of the regularized Gauss hypergeometric function~\cite[Eq. (07.24.26.0004.01)]{Mathematica}, we can express PD in closed form as in~\eqref{eq: Fox}, shown at the top of the next page, where $\mathcal{L}_{\textbf{s}}=\mathcal{L}_{\textbf{s}_1} \times \mathcal{L}_{\textbf{s}_2}$, and
\begin{align}
    \Phi  &= \frac{\Omega^{M-1} \exp \left[-  \Upsilon \ M \right]}{\Gamma (M-1)}\\ 
    \Omega & = \frac{M-1}{1-M+ \left( M-1\right){P_{\text{FA}}}^{1/(1-M)}}.
\end{align}
\normalsize
Observe that~\eqref{eq: Fox} has two new closed contours, $\mathcal{L}_{\textbf{s},1}$ and $\mathcal{L}_{\textbf{s},2}$. $\mathcal{L}_{\textbf{s},1}$ is an adjusted contour that appears due to the presence of the new gamma functions, whereas $\mathcal{L}_{\textbf{s},2}$ is the contour corresponding to the complex representation of the regularized Gauss hypergeometric function.
The integration paths for $\mathcal{L}_{\textbf{s},1}$ and $\mathcal{L}_{\textbf{s},2}$ are described in Section~\ref{sec:numerical_results}.
\begin{figure*}[!t]
\begin{flushleft}
\begin{align}
\label{eq: Fox}
P_{\text{D}} = \Phi \ \mathbf{H} \left[ \left[ \Omega,-  \Upsilon \ M \right];\left(\left[ 0,0,M-1,M\right],
\left(
    \begin{array}{c c c c}
     1 & 0 & -1  & -1 \\
     0 & 1 & 0 & -1 \\
    \end{array}
    \right)^T
\right);\left(\left[M,1\right], \left(
\begin{array}{cc}
 -1 & 0 \\
 0 & -1 \\
\end{array}
\right) \right); \mathcal{L}_{\textbf{s}} \right]
\end{align}
\normalsize
\end{flushleft}
\hrulefill
\end{figure*} 

A general implementation for the multivariate Fox's $H$-function is not yet available in mathematical packages such as MATHEMATICA, MATLAB, or MAPLE. 
Some works have been done to alleviate this problem~\cite{alhennawi16,yilmaz09,AlmeidaIEEE20}. Specifically in~\cite{alhennawi16}, the Fox's $H$-function was implemented from one up to four variables.
In this work, we provide an accurate and portable implementation in MATHEMATICA for the bivariate Fox's $H$-function.
The code used to compute~\eqref{eq: Fox} is presented in Appendix~D.
It is important to mention that such implementation is specific for our system model.
Moreover, an equivalent series  representation for~\eqref{eq: Fox} is also provided to facilitate the use of our results. This series representation is presented in the subsequent subsection.

\subsection{Infinite-Series Representation}
\label{sec: SeriesRepresentation}
Here, we provide a series representation for~\eqref{eq: Fox}.
To achieve this, we exploit the orthogonal selection of poles in Cauchy’s residue theorem.

First, let us consider the following suitable closed contours for \eqref{eq: Fox}: (i)~$\mathcal{L}_{\textbf{s},1}= \text{L}_{0,1}+\text{L}_{-\infty,1}$, and (ii) $\mathcal{L}_{\textbf{s},2}= \text{L}_{0,2}+\text{L}_{-\infty,2}$. Both contours are shown in Figs.~\ref{fig: L1} and~\ref{fig: L2}, where $\xi_1 \in \mathbb{R}^+$ must be chosen so that all the poles of $\Gamma(s_1)$ are separated from those of $\Gamma(M-1-s_1)$ and $\Gamma(M-s_1-s_2)$, and $\xi_2 \in \mathbb{R}^+$ must be chosen so that all the poles of $\Gamma(s_2)$  are separated from those of $\Gamma(M-s_1-s_2)$.
Additionally, $\rho_1$ and $\rho_2$ are the radius of the arcs $\text{L}_{-\infty,1}$ and $\text{L}_{-\infty,2}$, respectively. 

It is easy to prove that any complex integration along the paths $\text{L}_{-\infty,1}$ and $\text{L}_{-\infty,2}$ will be zero as $\rho_1$ and $\rho_2$ go to infinity, respectively. ($\rho_1$ and $\rho_2$ tend to infinity since the gamma functions $\Gamma (s_1)$ and $\Gamma (s_2)$ generate simple poles at all non-positive integers~\cite[Eq. (5.2.1)]{Olver10}.) 
Therefore, the final integration path for $\mathcal{L}_{\textbf{s},1}$ starts at $\xi_1 - j \infty$ and goes to $\xi_1 + j \infty$, whereas the final integration path for $\mathcal{L}_{\textbf{s},2}$ starts at $\xi_2 - j \infty$ and goes to $\xi_2 + j \infty$.

Now, we can rewrite~\eqref{eq: Fox} through the sum of residues as~\cite{Kreyszig10}
\begin{align}
    \label{eq:PD_series_1}
    P_{\text{D}} =\Phi \sum _{k=0}^{\infty } \sum _{l=0}^{\infty } \text{Res}\left[ \Xi \left(s_1,s_2 \right);s_1=-k,s_2=-l\right],
\end{align}
\normalsize
where $\text{Res} \left[ \Xi \left( s_1,s_2 \right);s_1-k,s_2=-l\right]$ represents the residue of $\Xi \left(s_1,s_2 \right)$ at the poles $s_1=-k$, $s_2=-l$, and
\begin{align}
\label{eq:integration_kernel}
\nonumber \Xi\left(s_1,s_2 \right)  = & \frac{\Gamma (s_1) \Gamma (s_2) \Gamma (M-s_1-1) \Gamma (-s_1+M-s_2)}{\Gamma (1-s_2) \Gamma (-(s_1-M))} \\
&\times \Omega^{-s_1}  \left(-  \Upsilon \ M \right)^{-s_2}.
\end{align}
\normalsize
is the integration kernel of~\eqref{eq: Fox}.

Accordingly, after applying the residue operation~\cite[Eq. (16.3.5)]{Kreyszig10},~\eqref{eq:PD_series_1} reduces to
\begin{align}
    \label{eq:eq:PD_series_2}
    \nonumber P_{\text{D}}= & \Phi \sum _{k=0}^{\infty } \sum _{l=0}^{\infty } \left\{ \frac{\Gamma (k+M-1) \Gamma (k+l+M)  \left(- \Omega \right)^k }{k! \Gamma (l+1)^2 \Gamma (k+M)} \right.\\
    & \times \left. \left( \Upsilon \ M\right)^l \right\}.
\end{align}
\normalsize
Finally, with the aid of~\cite[Eq. (15.2.1)]{Olver10} and after some mathematical manipulations, we obtain
\begin{align}
\label{eq:PD_series_3}
\nonumber P_{\text{D}}= &\exp \left[-  \Upsilon \ M \right] \Omega^{M-1}  \sum _{k=0}^{\infty }  \left\{ \frac{\Gamma (k+M) \left( \Upsilon \ M \right)^k }{\Gamma (k+1)^2} \right.\\
& \times \left.   \  _2\tilde{F}_1\left(M-1,k+M;M;-\Omega \right) \right\}.
\end{align}
\normalsize
It is worth mentioning that~\eqref{eq:PD_series_3} is also an original contribution of this work, proving to be very efficient and computationally tractable, as will be shown in the next section.

Generally, when radar designers need to compute the PD over a certain volume (i.e.,  range, azimuth and elevation), the calculation of the PD has to be performed for all the point scatterers within the entire coverage volume, thus increasing the computational load and simulation time.
Eq.~\eqref{eq:PD_series_3} can be executed quickly on an ordinary desktop computer, serving as a useful tool for radar designers.

Moreover, if $\mathcal{T}_0-1$ terms are used in~\eqref{eq:PD_series_3}, we can define the truncation error as
\begin{align}
    \label{eq:TE}
    \nonumber \mathcal{T}= & \frac{1}{\Gamma (M)}\sum _{k=T_0}^{\infty } \frac{\Omega ^{M-1} \exp \left[- M \Upsilon\right] (M \Upsilon)^k }{\Gamma (k+1)^2 } \\
    & \times \Gamma (k+M) \, _2F_1(M-1,k+M;M;\Omega ).
\end{align}
Since the Gauss hypergeometric function in (19) is monotonically decreasing with respect to $k$, $\mathcal{T}$ can be bounded as
\begin{align}
    \label{eq:TE2}
    \nonumber \mathcal{T}\leq & \, _2F_1\left(M-1,M+T_0;M;\Omega \right) \\
    & \times \sum _{k=T_0}^{\infty } \frac{\Omega ^{M-1} \exp \left[- M \Upsilon\right] (M \Upsilon)^k \Gamma (k+M)}{\Gamma (k+1)^2 \Gamma (M) }.
\end{align}
Since we add up strictly positive terms, we have
\begin{align}
    \label{eq:TE3}
    \nonumber & \sum _{k=T_0}^{\infty } \frac{\Omega ^{M-1} \exp \left[- M \Upsilon\right] (M \Upsilon)^k \Gamma (k+M)}{\Gamma (k+1)^2 \Gamma (M)} \\
    \nonumber & \ \ \ \leq \sum _{k=0}^{\infty } \frac{\Omega ^{M-1} \exp \left[- M \Upsilon\right] (M \Upsilon)^k \Gamma (k+M)}{\Gamma (k+1)^2 \Gamma (M)} \\
    &\ \ \ \overset{(a)}{=} \Omega ^{M-1} L_{M-1}(-M \Upsilon),
\end{align} 
where in step (a), we have used \cite[Eq. (05.02.02.0001.01)]{Mathematica} and some minor simplifications.
Then, from \eqref{eq:TE2} and \eqref{eq:TE3}, \eqref{eq:TE} can be bounded as 
\begin{align}
    \label{eq:UpperBound}
    \mathcal{T}\leq \frac{L_{M-1}(-M \Upsilon) \, _2F_1\left(M-1,M+T_0;M;-\Omega \right)}{\Omega ^{1-M}},
\end{align}
where $L_{\left(\cdot\right)}(\cdot)$ is the Laguerre polynomial~\cite[Eq. (05.02.02.0001.01)]{Mathematica}. 

\section{Numerical Results  and Discussions}
\label{sec:numerical_results}
\begin{figure}[t]
\begin{center}
\includegraphics[scale=0.43]{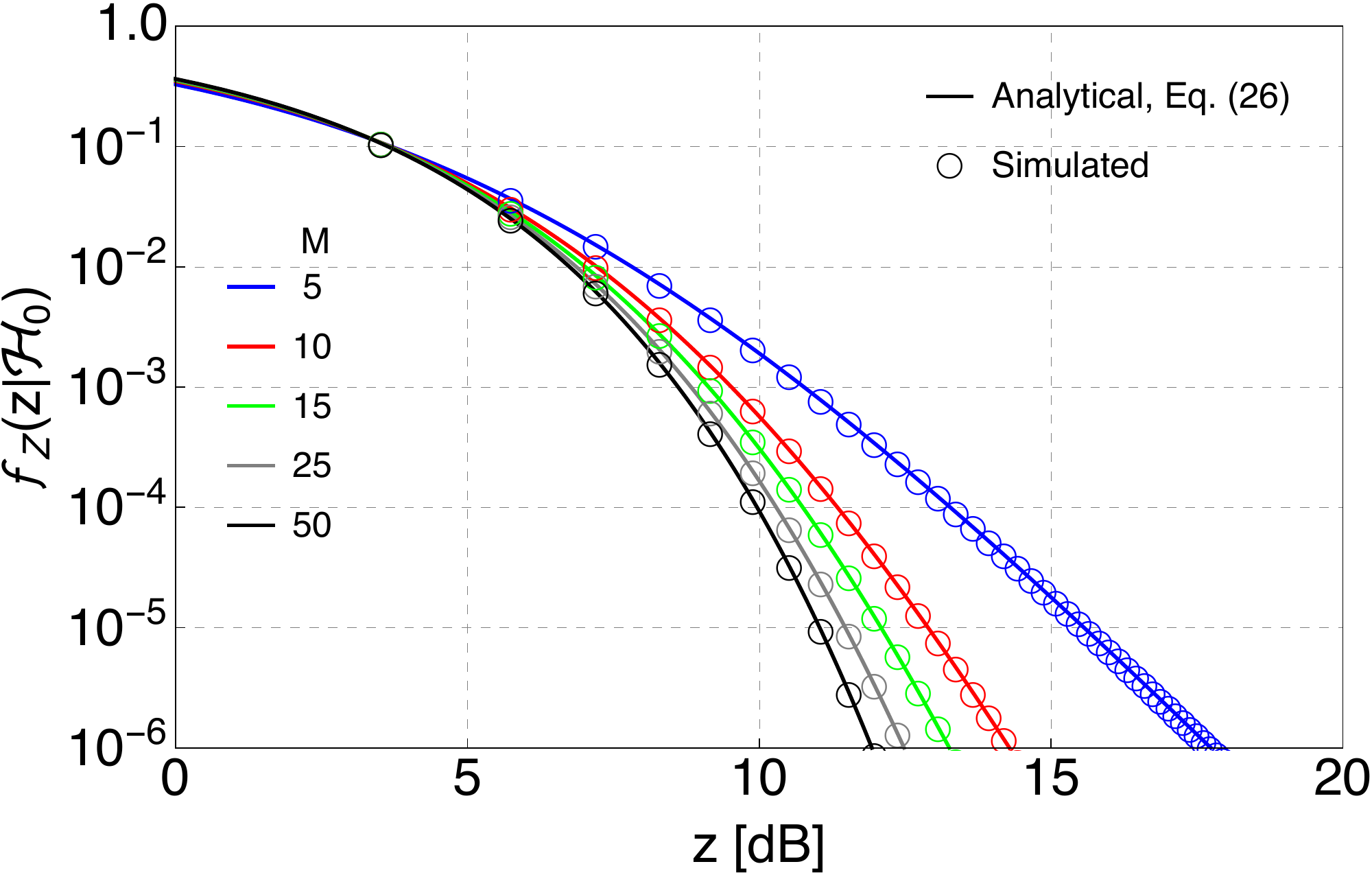}
\caption{PDF of $Z$ under $\mathcal{H}_0$ for different values of $M$.}
\label{fig:PdfH0}
\end{center} 
\end{figure}
\begin{figure}[t]
\begin{center}
\includegraphics[scale=0.43]{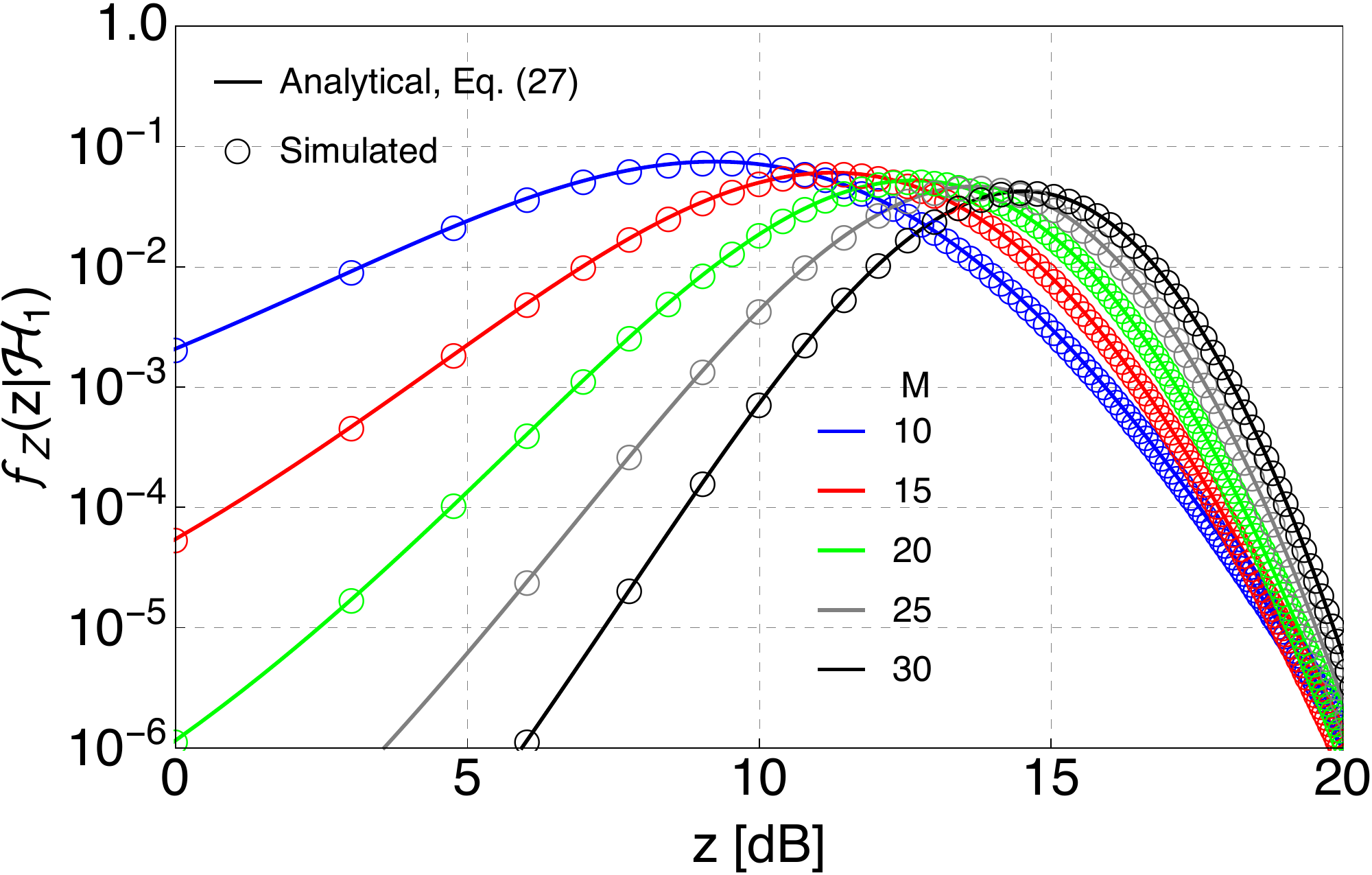}
\caption{PDF of $Z$ under $\mathcal{H}_1$ for different values of $M$ and $N$.}
\label{fig:PdfH1}
\end{center} 
\end{figure}
\begin{figure}[t]
\begin{center}
\includegraphics[scale=0.43]{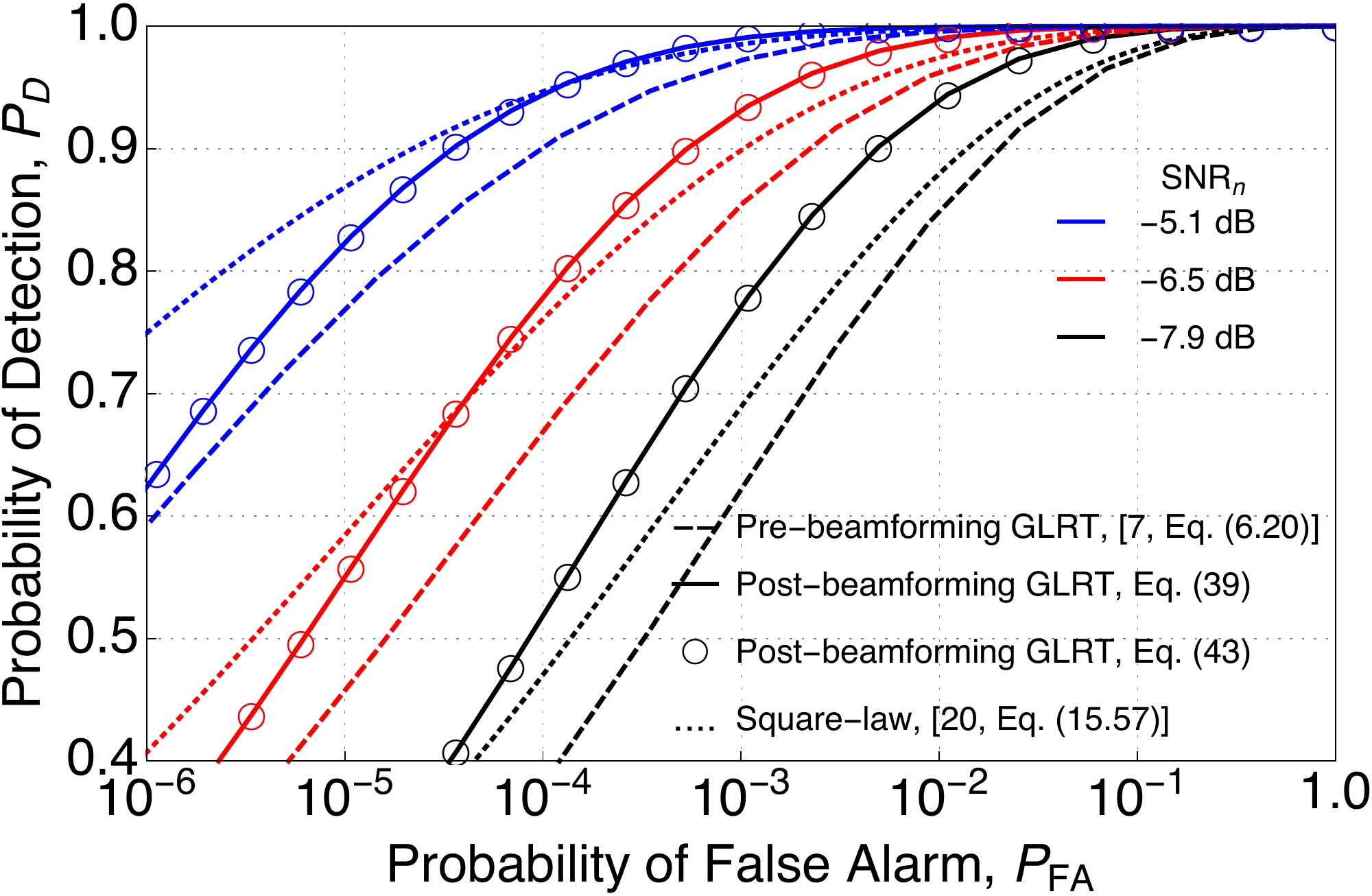}
\caption{$P_{\text{D}}$ vs $P_{\text{FA}}$ with $M=22$, $N=3$, and different values of $\text{SNR}_n$.}
\label{fig:PDvsPFA}
\end{center} 
\end{figure}
\begin{figure}[t]
\begin{center}
\includegraphics[scale=0.43]{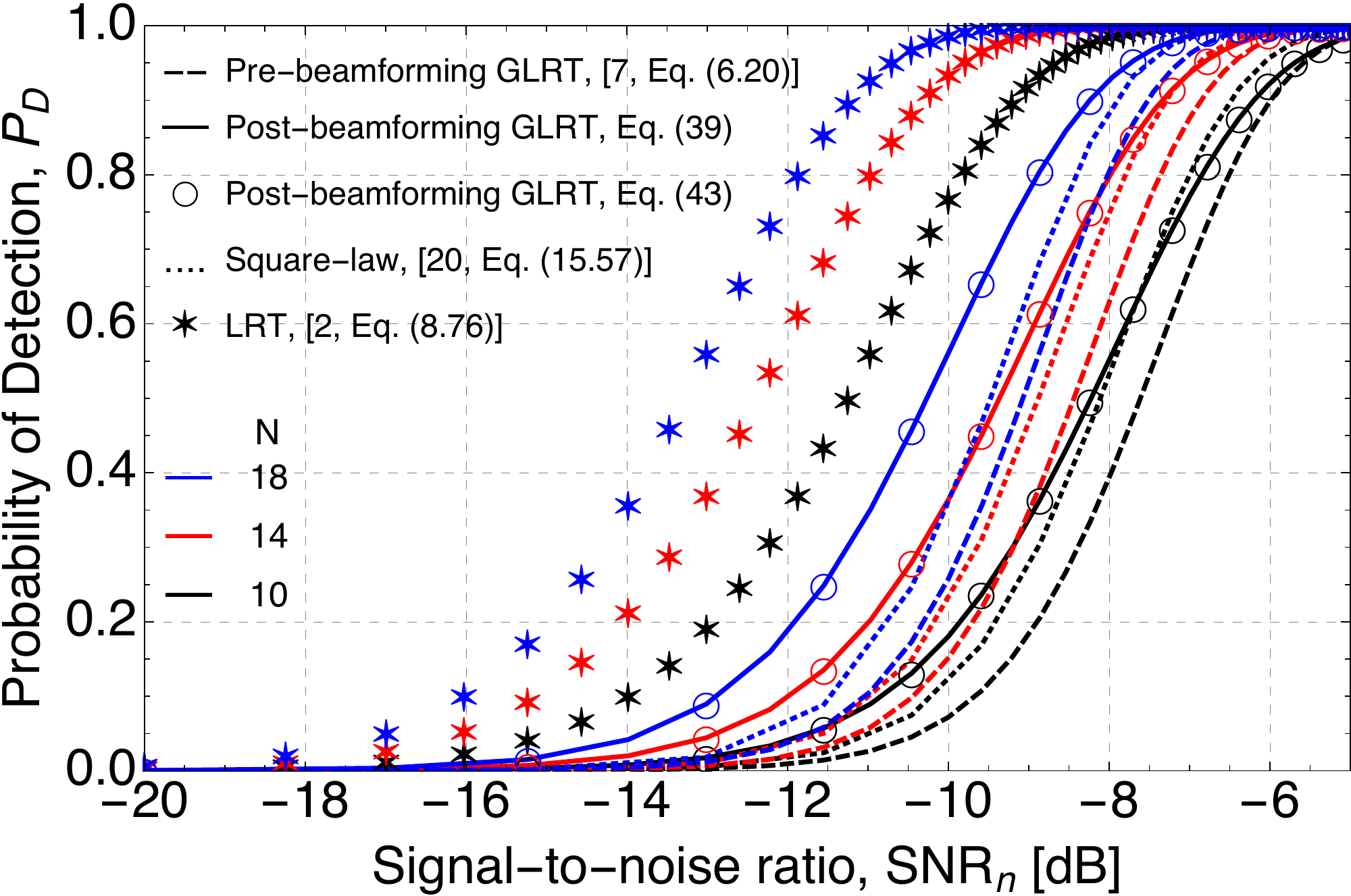}
\caption{$P_{\text{D}}$ vs $\text{SNR}_n$ with $M=15$, $P_{\text{FA}}=10^{-6}$ and different values of~$N$.}
\label{fig:PDvsSNR(Antennas)}
\end{center} 
\end{figure}
\begin{figure}[t]
\begin{center}
\includegraphics[scale=0.43]{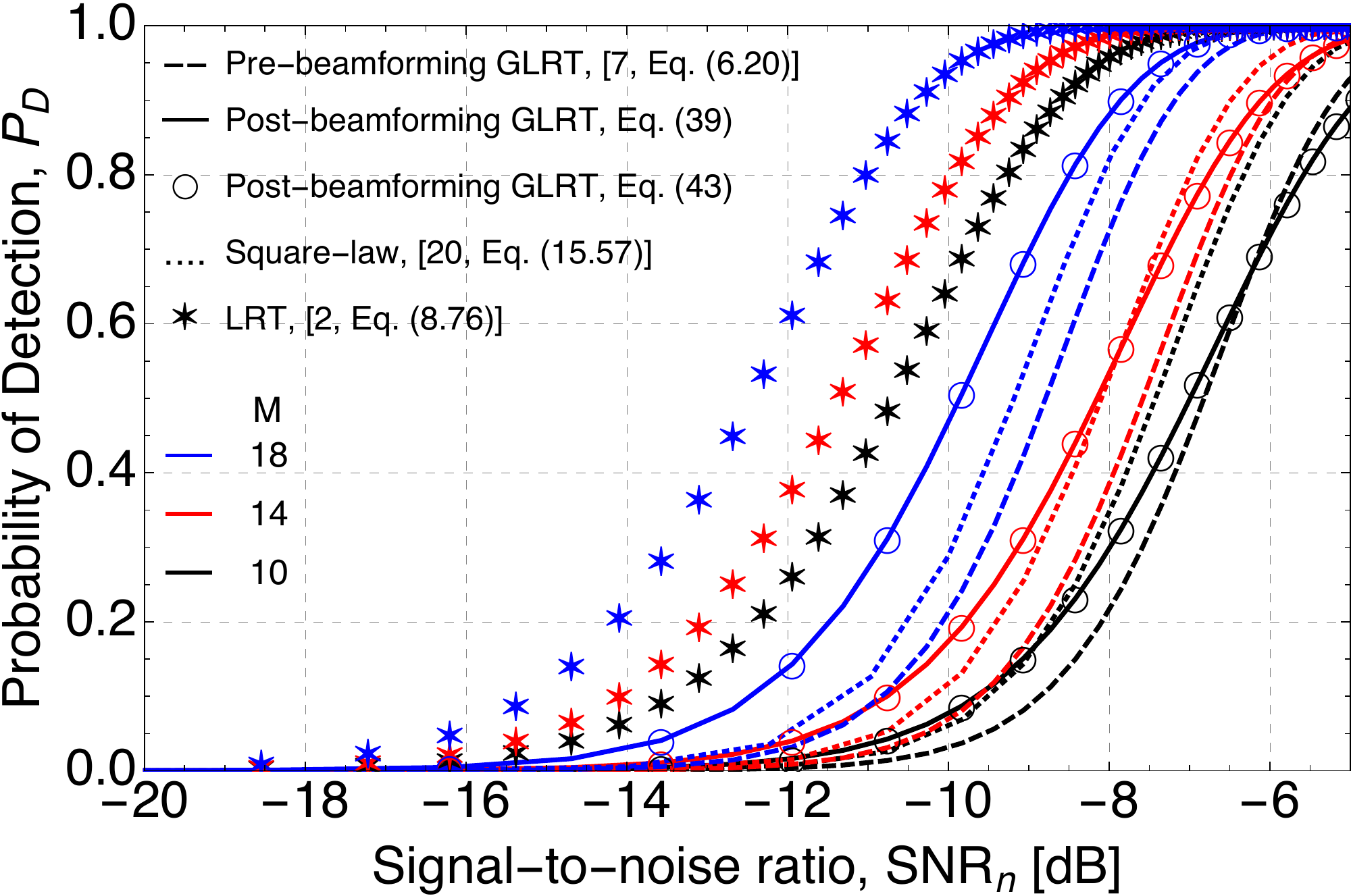}
\caption{$P_{\text{D}}$ vs $\text{SNR}_n$ with $N=11$, $P_{\text{FA}}=10^{-6}$ and different values of~$M$.}
\label{fig:SNRsamp}
\end{center} 
\end{figure}
\begin{figure}[t]
\begin{center}
\includegraphics[scale=0.43]{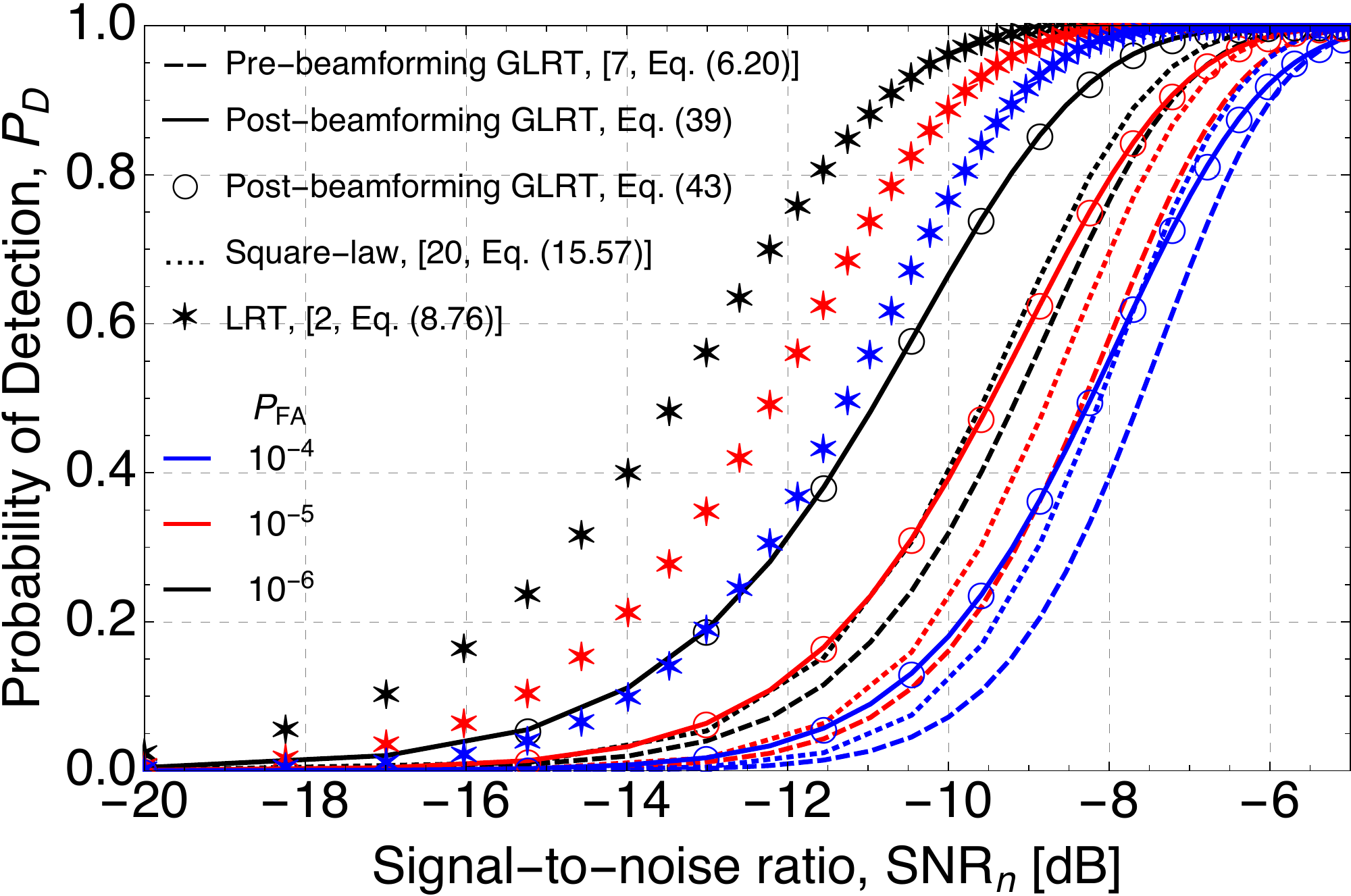}
\caption{$P_{\text{D}}$ vs $\text{SNR}_n$ with $M=10$, $N=15$ and different values of~$P_{\text{FA}}$.}
\label{fig:SNRpfa}
\end{center} 
\end{figure}
\begin{table*}[t]
\centering
\caption{Efficiency of~\eqref{eq:PD_series_3} as compared to \eqref{eq:PD_Integral_terms}.}
\label{tab: Table}
\begin{tabular}{ccccccc}
\hline \hline
$P_{\text{D}}$ Parameters  & $P_{\text{D}}$ Value &  \begin{tabular}[c]{@{}c@{}} Absolute  \\ Error,  $\epsilon$   \end{tabular}  & \begin{tabular}[c]{@{}c@{}}Number \\ of terms\end{tabular} & \begin{tabular}[c]{@{}c@{}}Computation Time \\ for Eq.~\eqref{eq:PD_Integral_terms} \end{tabular} &\begin{tabular}[c]{@{}c@{}}Computation Time \\ for Eq.~\eqref{eq:PD_series_3} \end{tabular}    & \begin{tabular}[c]{@{}c@{}}Reduction\\ Time\end{tabular}\\ \hline
$M=50$, $P_{FA}=10^{-8}$, $\Upsilon= -10 \ \text{dB}$ & $0.106$ \% & $5.471 \times 10^{-10}$ & 23 & $92.725 \times 10^{-3} \ \text{(s)}$  &$1.923 \times 10^{-3} \ \text{(s)}$   &  $97.92 \ \%$\\ \hline
$M=80$, $P_{FA}=10^{-8}$, $\Upsilon= -10 \ \text{dB}$ &  $1.416$ \%  & $5.248 \times 10^{-10}$ & 30 &$197.044 \times 10^{-3} \ \text{(s)}$  & $2.464 \times 10^{-3} \ \text{(s)}$ & $98.74 \ \%$ \\ \hline
$M=100$, $P_{FA}=10^{-8}$, $\Upsilon= -10 \ \text{dB}$ & $4.423$ \% & $6.032 \times 10^{-10}$& 34  &   $294.950 \times 10^{-3} \ \text{(s)}$  &$3.415 \times 10^{-3} \ \text{(s)}$   & $98.84 \ \%$ \\ \hline
$M=50$, $P_{FA}=10^{-8}$, $\Upsilon= -5 \ \text{dB}$ & $19.224$ \% & $5.261 \times 10^{-10}$& 45  & $96.370 \times 10^{-3} \ \text{(s)}$ &$4.625  \times 10^{-3} \ \text{(s)}$  &  $95.20 \ \%$ \\
\hline
$M=50$, $P_{FA}=10^{-6}$, $\Upsilon= -5 \ \text{dB}$ & $52.886$ \% & $5.341 \times 10^{-10}$& 45  & $95.769 \times 10^{-3} \ \text{(s)}$ &$4.663 \times 10^{-3} \ \text{(s)}$  &  $95.13 \ \%$ \\
\hline
$M=50$, $P_{FA}=10^{-4}$, $\Upsilon=- 5 \ \text{dB}$  & $87.958$ \% &$5.361 \times 10^{-10}$ & 45  & $92.911 \times 10^{-3} \ \text{(s)}$ & $4.54 \times 10^{-3} \ \text{(s)}$  & $95.11 \ \%$ \\ \hline
$M=50$, $P_{FA}=10^{-6}$, $\Upsilon= -3 \ \text{dB}$ & $92.089$ \% &$9.339 \times 10^{-10}$ & 60  & $99.896 \times 10^{-3} \ \text{(s)}$  & $7.043 \times 10^{-3} \ \text{(s)}$ &  $92.94 \ \%$ \\
\hline
$M=50$, $P_{FA}=10^{-6}$, $\Upsilon= -2 \ \text{dB}$  & $98.621$ \% &$4.790 \times 10^{-10}$ & 71  & $95.124 \times 10^{-3} \ \text{(s)}$ & $9.238 \times 10^{-3} \ \text{(s)}$ & $90.28 \ \%$ \\ \hline
$M=50$, $P_{FA}=10^{-6}$, $\Upsilon= -1 \ \text{dB}$ & $99.902$ \% &$6.522 \times 10^{-10}$ & 83  & $98.728  \times 10^{-3} \ \text{(s)}$   & $11.418 \times 10^{-3} \ \text{(s)}$ & $88.43 \ \%$ \\
\hline  \hline
\end{tabular}
\end{table*}
In this section, we validate our derived expressions and discuss the representative results. 
To do so, we make use of the~\textit{receiver operating characteristic} (ROC) curves and Monte-Carlo simulations.\footnote{The number of realizations was set to $1 \times 10^{7}$.}
For comparison purposes, besides the pre-beamforming GLRT and square-law detectors, we also include the (optimum) LRT detector~\cite{kay98} so as to quantify the SNR losses.\footnote{Herein, the SNR loss is defined as extra SNR required to achieved the same performance as the LRT detector~\cite[Eq. (4.3)]{kay98}, for a given PD.}
 
Figs.~\ref{fig:PdfH0} and~\ref{fig:PdfH1} show the PDF of $Z$ (analytical and simulated) given the hypotheses $\mathcal{H}_0$ and $\mathcal{H}_1$, respectively. The distribution parameters have been selected to show the broad range of shapes that the PDFs can exhibit. 
Observe the perfect match between Monte-Carlo simulations and our derived expressions [refer to~\eqref{eq:PDF_h0} and \eqref{eq:pdfH1_4}].

Fig.~\ref{fig:PDvsPFA} shows $P_{\text{D}}$ as a function of $P_{\text{FA}}$ (analytical and simulated) for different values of $\text{SNR}_n$. 
Observe that for low $\text{SNR}_n$, the \post{} is superior to both the \pre{} and the square-law detector. That is, the weaker the signals, the better the performance of our proposed detector.
For example, given $P_{\text{FA}}=10^{-4}$, the \post{}, the \pre{}, and the square-law detector provide, respectively, the following probabilities of detection: $0.53$, $0.38$ and $0.47$ for $\text{SNR}_n=-7.9$~dB; $0.78$, $0.66$ and $0.75$ for $\text{SNR}_n=-6.5$ dB; and finally, $0.94$, $0.90$ and $0.95$ for $\text{SNR}_n=-5.1$ dB. 
The following figures illustrate the impact on the PD as the SNR is reduced. 

Fig.~\ref{fig:PDvsSNR(Antennas)} shows $P_{\text{D}}$ as a function of $\text{SNR}_n$ (analytical and simulated) for different values of $N$. Note that all detectors improve as the number of antennas increases, requiring a lower SNR for a certain PD. Also, note how the \post{} overcomes the \pre{} and the square-law detector as the SNR decreases.
For example, given $\text{SNR}_n=-8$ dB, the \post{}, the \pre{}, and the square-law detector provide, respectively, the following probabilities of detection: $0.55$, $0.40$ and $0.54$ for $N=10$; $0.79$, $0.64$ and $0.75$ for $N=14$; and finally, $0.94$, $0.80$ and $0.86$ for $N=18$.
Additionally, observe how the SNR loss is reduced as $N$ increases. 
In particular, for a fixed $P_{\text{D}}=0.8$, the \post{} is superior to both the \pre{} and the square-law detector deliver, respectively, the following SNR losses: $3.8$ dB, $4.2$ dB and $2.8$ dB for $N=10$; $2.9$ dB, $3.6$ dB and $3.1$ dB for $N=14$; and finally, $2.8$ dB, $3.9$ dB and $3.5$ dB for $N=18$.

Fig.~\ref{fig:SNRsamp} shows $P_{\text{D}}$ as a function of $\text{SNR}_n$ (analytical and simulated) for different values of $M$. Observe that all detectors improve as the number of samples increases.
This occurs because we ``average down'' the noise power by increasing $M$. Once again, the \post{} performs better than the \pre{} and the square-law detector in the low SNR regime.
More specifically, given $\text{SNR}_n=-8$ dB, the \post{}, the \pre{} and the square-law detector provide, respectively, the following probabilities of detection: $0.30$, $0.21$ and $0.35$ for $M=10$; $0.53$, $0.40$ and $0.53$ for $M=14$; and finally, $0.87$, $0.73$ and $0.82$ for $M=18$.
Moreover, observe how the SNR loss is reduced as $N$ increases.  
In particular, for a fixed $P_{\text{D}}=0.8$, the \post{}, the \pre{} and the square-law detector deliver, respectively, the following SNR losses: $3.6$ dB, $3.4$ dB and $3.2$ dB for $M=10$; $3.4$ dB, $3.5$ dB and $3.1$ dB for $M=14$; and finally, $2.8$ dB, $3.6$ dB and $3.1$ dB for $M=18$.

Fig.~\ref{fig:SNRpfa} shows $P_{\text{D}}$ as a function of $\text{SNR}_n$ (analytical and simulated) for different values of $P_{\text{FA}}$. Note that all detectors improve as $P_{\text{FA}}$ is increased.  
This fundamental trade-off means that if the PFA is reduced, the PD decreases as well. Observe that for low SNR, the superiority of our detector still remains.
For example, given $\text{SNR}_n=-8$ dB, the \post{}, the \pre{} and the square-law detector provide, respectively, the following probabilities of detection: $0.93$, $0.76$ and $0.84$ for $P_{\text{FA}}=10^{-6}$; $0.80$, $0.57$ and $0.70$ for $P_{\text{FA}}=10^{-5}$; and finally, $0.55$, $0.40$ and $0.54$ for $P_{\text{FA}}=10^{-4}$.
Additionally, observe how the SNR loss is reduced as $N$ increases.  
In particular, for a fixed $P_{\text{D}}=0.8$, the \post{}, the \pre{} and the square-law detector deliver, respectively, the following SNR losses: $2.4$ dB, $3.6$ dB and $3.2$ dB for $P_{\text{FA}}=10^{-6}$; $2.6$ dB, $3.4$ dB and $3.0$ dB for $P_{\text{FA}}=10^{-5}$; and finally, $2.9$ dB, $3.2$ dB and $2.8$ dB for $P_{\text{FA}}=10^{-4}$.

An important remark is in order. The results presented herein show that if the the received signals are weak, instead of processing the received signals separately, as described in~\cite[Eq. (6.20)]{kay98}, it is better to sum up the signals and then construct the system's \textit{detection statistic}.
Intuitively, this means that if the signal received by each antenna is defectively estimated (due to low target power or strong interference), then the system will also deliver a faulty final estimate.
Therefore, it is better to reinforce (i.e., applying the beamforming operation) the overall signal before any further pre-processing.
Moreover, the way we create the system's \textit{detection statistic} enables us to improve radar detection as we increase the number of antennas while maintaining a fixed PFA.

Table~\ref{tab: Table} illustrates the efficiency of~\eqref{eq:PD_series_3} by showing the absolute error, computation time, required number of terms to guarantee a certain accuracy, and reduction time [compared to \eqref{eq:PD_Integral_terms}]. The absolute error can be expressed~as
\begin{align}
    \label{}
    \epsilon = | P_{\text{D}} - \overline{P_{\text{D}}}|,
\end{align}
where $\overline{P_{\text{D}}}$ is the probability of detection obtained via MATHEMATICA's built-in numerical integration.\footnote{Eq.~\eqref{eq:PD_Integral_terms} was evaluated by using the fastest MATHEMATICA's integration method, ``\texttt{GlobalAdaptive}'', with an accuracy goal of $10^{-10}$.} 
Observe that for 9 different parameter settings,~\eqref{eq:PD_series_3} converges rapidly requiring between 23 and 83 terms to guarantee an accuracy of $10^{-10}$. 
Moreover, the computation time dropped dramatically, thereby providing reduction times above $88$\%.
This impressive reduction can lead to major savings in computational load if one wants to evaluate the detection performance over an entire area or volume covered by the radar system.

\section{Conclusions}
\label{sec:conclusions}
This paper proposed and analyzed a new GLRT phased array detector, which is projected after the analog beamforming operation. 
For the analysis, a \textit{nonfluctuating} target embedded in CWGN was considered.
From the practical point of view, this detector fulfils the hardware and computational constraints of most radar systems.
The performance metrics -- PD and PFA -- were derived in \textit{closed form} assuming a total lack of knowledge about the target echo and noise statistics. 
Moreover, a novel fast converging series for the PD was also derived.  
This series representation proved to be very efficient and computationally tractable, showing an outstanding accuracy and impressive reductions in both computational load and computation time, compared to MATHEMATICA’s built-in numerical integration.
Numerical results showed that when the incoming signals are weak, it is best to combine (sum) them before any estimation or further processing.
Indeed, this paper is conclusive in indicating that for low SNR, the \post{} shows superior to the \pre{} and square-law detectors.
Another interesting feature about the \post{} demonstrates that for a fixed PFA, the detection threshold is independent of the number of antennas, which allows us to improve the PD (by increasing $N$) while maintaining a fixed PFA.
The SNR losses were also quantified and they illustrated the superiority of the \post{} as $N$ and $M$ increase.

\section*{Appendix A: Proof of Lemma 1}
\label{app:lemma1}
Let us define the following RV
\begin{align}
    \label{eq:I1}
    \mathcal{I}_3 \triangleq \frac{1}{N \sigma^2}  \overset{M}{\sum _{m=1}} \left( \mathbf{Re} \left[ r_m\right] - \mu_X \right)^2,
\end{align}
where $\mu_X$ is the total sum of the target echoes for the in-phase components. 

Rewriting ~\eqref{eq:I1}, we have
\begin{align}
    \label{eq:I1_V2}
    \mathcal{I}_3=\overset{M}{\sum _{m=1}} \left( \frac{\mathbf{Re} \left[ r_m\right]- \mu_X }{\sqrt{N} \sigma}\right)^2.
\end{align}
\normalsize
It can be noticed that $\mathcal{I}_3$ is a sum of the squares of $M$  standard Gaussian (zero mean and unit variance) RVs. Therefore, $\mathcal{I}_3$ can be modeled by a CCS RV with $M$ degrees of freedom.

Now, after performing some manipulations, we can rewrite~\eqref{eq:I1_V2} as 
\begin{align}
    \label{eq:I1expand}
    \nonumber \mathcal{I}_3  = & \overset{M}{\sum _{m=1}} \left( \frac{\mathbf{Re} \left[ r_m\right]- \hat{\mu}_X }{\sqrt{N} \sigma} + \frac{\hat{\mu}_X- \mu_X }{\sqrt{N} \sigma}\right)^2 \\
    \nonumber  \overset{(a)}{=} & \overset{M}{\sum _{m=1}} \left( \frac{\mathbf{Re} \left[ r_m\right]- \hat{\mu}_X }{\sqrt{N} \sigma} \right)^2 + 2 \left( \frac{\hat{\mu}_X- \mu_X }{\sqrt{N} \sigma} \right) \\
    \nonumber & \times  \left(\frac{\sum^{M}_{m=1} \mathbf{Re} \left[ r_m\right]- M \hat{\mu}_X }{\sqrt{N} \sigma} \right) + \overset{M}{\sum _{m=1}} \left( \frac{\hat{\mu}_X- \mu_X }{\sqrt{N} \sigma}\right)^2 \\
    \overset{(b)}{=} & \underbrace{\overset{M}{\sum _{m=1}} \left( \frac{\mathbf{Re} \left[ r_m\right]- \hat{\mu}_X }{\sqrt{N} \sigma} \right)^2 }_{\triangleq \ \mathcal{I}_{4}}  + \underbrace{\left( \frac{\hat{\mu}_X- \mu_X }{\sqrt{N} \sigma / M}\right)^2}_{\triangleq \ \mathcal{I}_{5}},
\end{align}
where in step (b) we use the fact that $M \hat{\mu}_X = \sum^{M}_{m=1} \mathbf{Re} \left[ r_m\right]$ and, consequently, the second term in step~(a) vanishes.
Observe that $\mathcal{I}_5$ represents the square of a standard Gaussian variable and, therefore, can be modeled by a CCS distribution with one degree of freedom. 

Employing the additivity property of the CCS distribution~\cite{papoulis02} and taking into account the distributions of $\mathcal{I}_3$ and $\mathcal{I}_5$, we can now describe $\mathcal{I}_4$ by a CCS RV with $M-1$ degrees of freedom.
Also, observe that $\mathcal{I}_4$ is just the first term of~\eqref{eq:I2_final}. 

Following the same approach, it can be prove that the second term in~\eqref{eq:I2_final} also follows a CCS distribution with $M-1$ degrees of freedom. Since $\mathcal{I}_2$ is formed by the sum of two CCS RVs, then its distribution is governed by a CCS RV with $2(M-1)$ degrees of freedom, which completes the proof.
It is worth mentioning that this result remains true regardless of the hypothesis, because any value of $\mu_X$ or $\mu_Y$ will not affect the distribution of $\mathcal{I}_2$. 

\section*{Appendix B: Proof of Lemma 2}
\label{app:lemma2}
Let
\begin{align}
    \label{eq:P1}
    P_1&=\mathbf{L} \left(\mathbf{L}^T \mathbf{L} \right)^{-1}\mathbf{L}^T=\frac{1}{M}\mathbf{L} \  \mathbf{L}^T \\ \label{eq:P2}
    P_2&=\mathbf{I}-P_1=\mathbf{I}-\frac{1}{M}\mathbf{L} \  \mathbf{L}^T
\end{align}
\normalsize
be symmetric and idempotent matrices such that $\text{rank} \left( P_1 \right)=\mathbf{L}$, $\text{rank} \left( P_2 \right)=M-1$ and  $P_1+P_2=\mathbf{I}$, where $\mathbf{I}\in \mathbb{N}^{M \times M}$ represents the identity matrix and $ \mathbf{L}=\left[1,1,\cdots,1 \right]^T \in \mathbb{N}^M$ is the unitary vector.
In addition, let 
\begin{align}
\label{eq:Xvector}
\mathbf{Re} \left[ \underline{r}\right]=\left[\mathbf{Re} \left[ r_1\right],\mathbf{Re} \left[ r_2\right],\cdots,\mathbf{Re} \left[ r_M\right]\right]^T
\end{align}
be a random vector with $\mathbb{E}\left[\mathbf{Re} \left[ \underline{r}\right] \right]= \mu_X \mathbf{L}$ and $\mathbb{COV}\left[ \mathbf{Re} \left[ \underline{r}\right] \right]=N \sigma ^2 \mathbf{I} $. Then, the Cochran's Theorem~\cite{cochran34} states that
\begin{align}
    \label{eq:a1}
    \omega_1=&\frac{\mathbf{Re} \left[ \underline{r}\right]^T P_1 \  \mathbf{Re} \left[ \underline{r}\right]}{N \sigma^2}\\ \label{eq:a2}
    \omega_2=&\frac{\mathbf{Re} \left[ \underline{r}\right]^T P_2 \  \mathbf{Re} \left[ \underline{r}\right]}{N \sigma^2}
\end{align}
\normalsize
are independently distributed.

\noindent
Now, replacing~\eqref{eq:P1} in~\eqref{eq:a1}, we have
\begin{align} 
    \nonumber \omega_1&=\frac{1}{N \sigma ^2} \mathbf{Re} \left[ \underline{r}\right]^T \left(\frac{1}{M} \mathbf{L} \  \mathbf{L}^T \right) \mathbf{Re} \left[ \underline{r}\right]\\
    \nonumber &= \frac{1}{M N \sigma ^2}\mathbf{Re} \left[ \underline{r}\right]^T \mathbf{L} \  \mathbf{L}^T \mathbf{Re} \left[ \underline{r}\right]\\  \label{eq:ProveP1}
    &=\frac{1}{M N \sigma ^2} \left( \sum _{k=1}^M \mathbf{Re} \left[ r_k\right] \right)^2.
\end{align}
\normalsize
Similarly, inserting~\eqref{eq:P2} in~\eqref{eq:a2}, we have
\begin{align} 
    \nonumber \omega_2 & \overset{(a)}{=} \frac{1}{N \sigma ^2}\mathbf{Re} \left[ \underline{r}\right]^T  P_2^T P_2   \mathbf{Re} \left[ \underline{r}\right]\\
    \nonumber &= \frac{1}{N \sigma ^2} \left\| P_2 \mathbf{Re} \left[ \underline{r}\right] \right\|^2\\
    \nonumber &\overset{(b)}{=} \frac{1}{N \sigma ^2} \left\| \left(\mathbf{I}-\frac{1}{M}\mathbf{L} \ \mathbf{L}^T \right)\mathbf{Re} \left[ \underline{r}\right] \right\|^2\\
    \nonumber &\overset{(c)}{=}\frac{1}{N \sigma ^2} \left\| \mathbf{Re} \left[ \underline{r}\right] -  \mathbf{L}\hat{\mu}_X \right\|^2\\   \label{eq:ProveP2}
    &\overset{(d)}{=}\frac{1}{N \sigma ^2} \sum _{k=1}^M \left( \mathbf{Re} \left[ r_k\right] - \hat{\mu}_X \right)^2,
\end{align}
\normalsize
where in step (a), we have used the definition of idempotent and symmetric matrices~\cite{springer79}, in step (b), we have used~\eqref{eq:P2}, in step (c), we have employed~\eqref{eq:MLE_AX}, and in step (d), we have used~\eqref{eq:Xvector} and applied the Euclidean norm.
Observe that $\omega_1$ and $\omega_2$ are the first terms of~\eqref{eq:I1_final} and~\eqref{eq:I2_final}, respectively. The same approach can also be applied to prove the independence between the second terms.
Finally, since $\mathbf{Re} \left[ r_k\right]$ and $\mathbf{Im} \left[ r_k\right]$ are also independent statistics (cf. Section~\ref{sec:HypotesisTest}), then $\mathcal{I}_1$ and $\mathcal{I}_2$ are mutually independent RVs, which completes the proof.

\section*{Appendix C: Derivation of \eqref{eq:pdfH1_4}}
\label{app:2}
To prove~\eqref{eq:pdfH1_4}, we make use of the doubly noncentral F-distribution, defined as~\cite{Bulgren71}
\begin{align}
\label{eq:Doubly_F}
\nonumber & \mathit{f}_Z  \left(z| \mathcal{H}_1\right)=\sum _{k=0}^{\infty } \sum _{l=0}^{\infty } \left\{ \frac{z^{-1} \exp \left[\frac{-\lambda _1-\lambda _2}{2} \right] \left(\frac{\alpha_1 z}{\alpha_1 z+\alpha_2}\right){}^{\frac{\alpha_1}{2}}}{k! \ l! \ B\left(k+\frac{\alpha_1}{2},l+\frac{\alpha_2}{2}\right)} \right. \\
& \left. \times \left(\frac{\alpha_2}{\alpha_1 z+\alpha_2}\right)^{\frac{\alpha_2}{2}} \left(\frac{\lambda _1 \alpha_1 z}{2 \left(\alpha_1 z+\alpha_2\right)}\right)^k \left(\frac{\lambda _2 \alpha_2}{2 \left(\alpha_1 z+\alpha_2\right)}\right)^l \right\}
\end{align}
\normalsize
Rearranging some terms, and after applying~\cite[Eq. (07.20.02.0001.01)]{Mathematica},~\eqref{eq:Doubly_F} simplifies to
\begin{align}
\label{eq:PDF_h1_v1}
\nonumber \mathit{f}_Z & \left(z| \mathcal{H}_1\right) = z^{-1} \exp \left[\frac{-\lambda _1-\lambda _2}{2} \right] \left(\frac{\alpha_1 z}{\alpha_1 z+\alpha_2}\right)^{\frac{\alpha_1}{2}}\\
\nonumber & \times  \left(\frac{\alpha_2}{\alpha_1 z+\alpha_2}\right)^{\frac{\alpha_2}{2}} \sum _{k=0}^{\infty } \left\{ \left(\frac{\lambda _1 \alpha_1 z}{2 \alpha_1 z+2 \alpha_2}\right)^k  \right.\\
& \times \left.  \frac{_1F_1\left(\frac{1}{2} \left(2 k+\alpha_1+\alpha_2\right);\frac{\alpha_2}{2};\frac{\alpha_2 \lambda _2}{2 \left(z \alpha_1+\alpha_2\right)}\right) }{k! \  B\left(k+\frac{\alpha_1}{2},\frac{\alpha_2}{2}\right)} \right\}.
\end{align} 
\normalsize
Now, replacing $\alpha_1=2$, $\alpha_2=2(M-1)$, $\lambda_1=M(\mu_X^2+\mu_Y^2)/N \sigma^2$, and $\lambda_2=0$ (cf. Section~\ref{sec:Decision}) in~\eqref{eq:PDF_h1_v1}, and after applying~\cite[Eq. (15.2.1)]{Olver10}, and~\cite[Eq. (5.12.1)]{Olver10}, we obtain 
\begin{align}
\label{eq:PDF_h1_v2}
\nonumber \mathit{f}_Z  \left(z| \mathcal{H}_1\right) = & \frac{ \exp \left[-\frac{M \left(\mu_{X}^2+\mu_{Y}^2\right)}{2 N\sigma^2} \right]}{\Gamma (M)} \left(\frac{M-1}{M+z-1}\right)^M\\
& \times \sum _{k=0}^{\infty }\frac{\Gamma (k+M) }{\Gamma (k+1)^2} \left(\frac{M  z \left(\mu_{X}^2+\mu_{Y}^2\right)}{2 N \sigma^2 (M+z-1)}\right)^k.
\end{align}
\normalsize
Finally, after using the definition of the Kummer confluent hypergeometric function~\cite[Eq. (07.20.02.0001.01)]{Mathematica}, along with minor simplifications, we obtain~\eqref{eq:pdfH1_4}, which completes the derivation. 

\section*{Appendix D: Mathematica's implementation for the Bivariate Fox's $H$-function}
\label{sec:Appendix D}
\lstset{
	tabsize=4,
	frame=single,
	language=mathematica,
	basicstyle=\scriptsize\ttfamily,
	keywordstyle=\color{black},
	backgroundcolor=\color{white},
	commentstyle=\color{magenta},
	showstringspaces=false,
	emph={
		Newton,Newton_, x_, delta_, delta, D_, D,beta_,beta, B_, M_, M, L_, L, s, s_,B
	},emphstyle={\color{olive}},
	emph={[2] H, ClearAll, Remove, H_, Ns, value
	},emphstyle={[2]\color{blue}},
	emph={[3] UpP, LoP, Theta, R1, T1, R2, T2, k, R1, T1, R2, T2, m, n,kernel, S, W, t},emphstyle={[3]\color{magenta}}
	}
\vspace{0.5cm}
\begin{CenteredBox}
\begin{lstlisting}[caption={},linewidth=8cm, label=code:label]
ClearAll["Global`*"]; Remove[s];
H[x_, delta_, D_,beta_, B_]
 := Module[{UpP, LoP, Theta,R1, T1, R2, T2, m, n},
 L=Length[Transpose[D]]; (*L represents the
 dimension of the Fox's H-function*)
 m=Length[D]; (*Number of Gamma functions in the
 numerator*)
 n=Length[B]; (*Number of Gamma functions in the
 denominator*)
 S=Table[Subscript[s,i],{i,1,L}]; (*s is the 
 vector containing the number of branches, in our 
 case s=[s_1,s_2]*)
 UpP=Product[Gamma[delta[[1,j]]+Sum[D[[j,k]]
     S[[k]],{k,1, L}]], {j,1,m}];
 LoP=Product[Gamma[beta[[1,j]]+Sum[B[[j,k]]
     S[[k]],{k,1,L}]],{j,1,n}];
 Theta=UpP/LoP (*Theta computes Eq. (2)*);
 W=50; (*Limit for the complex integration*)
  T=Table[delta[[1,j]]+Sum[D[[j,k]]
  S[[k]],{k,1,L}]>0,{j,1,m}]; (*Generates
  a restriction table*)
 R1=Reduce[And@@Flatten[{T[[1]],T[[3]]}]];
 (*R1 computes the real interval that separates 
 the poles of Gamma[s_1] from the poles of
 Gamma[M-1-s_1] and Gamma[M-s_1-s_2]*)
 T1=Mean[{First@R1,Last@R1}]; 
 R2=Reduce[And@@Flatten[{T[[2]],T[[4]]}]];
 (*R2 computes the real interval that separates
 the poles of Gamma[s_2] from the poles of
 Gamma[M-s_1-s_2]*)
 T2=Mean[{First@R2,Last@R2}]; 
 W=100; (*Limit for the complex axis*)
 kernel=Theta(x[[1,1]])^(-S[[1]])(x[[1,2]])^(-S[[2]])
   /.{S[[1]]->s1,S[[2]]->s2}; (*Prepare the Kernel 
   for Mathematica's Integration*)
 N[1/(2*Pi*I)^2 NIntegrate[kernel,{s1,T1-I W,T1+I W},
 {s2,T2-I W,T2+I W}],20]]
\end{lstlisting}
\end{CenteredBox}

\bibliographystyle{IEEEtran}
\bibliography{Bibliography}

\begin{thebibliography}{10}
\providecommand{\url}[1]{#1}
\csname url@samestyle\endcsname
\providecommand{\newblock}{\relax}
\providecommand{\bibinfo}[2]{#2}
\providecommand{\BIBentrySTDinterwordspacing}{\spaceskip=0pt\relax}
\providecommand{\BIBentryALTinterwordstretchfactor}{4}
\providecommand{\BIBentryALTinterwordspacing}{\spaceskip=\fontdimen2\font plus
\BIBentryALTinterwordstretchfactor\fontdimen3\font minus
  \fontdimen4\font\relax}
\providecommand{\BIBforeignlanguage}[2]{{%
\expandafter\ifx\csname l@#1\endcsname\relax
\typeout{** WARNING: IEEEtran.bst: No hyphenation pattern has been}%
\typeout{** loaded for the language `#1'. Using the pattern for}%
\typeout{** the default language instead.}%
\else
\language=\csname l@#1\endcsname
\fi
#2}}
\providecommand{\BIBdecl}{\relax}
\BIBdecl

\bibitem{blake86}
L.~V. Blake, \emph{Radar Range-performance Analysis}, 1st~ed.\hskip 1em plus
  0.5em minus 0.4em\relax Norwood, MA, USA: Artech House, 1986.

\bibitem{leon94}
A.~Leon-Garcia, \emph{Probability and Random Processes for Electrical
  Engineering}, 3rd~ed.\hskip 1em plus 0.5em minus 0.4em\relax New Jersey, NJ,
  USA: Pearson Prentice Hall, 1994.

\bibitem{chernoff54}
H.~Chernoff, ``On the distribution of likelihood ratio,'' \emph{Ann. Math.
  Statist.}, vol.~25, no.~3, pp. 573--578, Sept. 1954.

\bibitem{kay93}
S.~M. Kay, \emph{Fundamentals of Statistical Signal Processing: Estimation
  Theory}, 1st~ed.\hskip 1em plus 0.5em minus 0.4em\relax Upper Saddle River,
  NJ, USA: Prentice Hall PTR, 1993.

\bibitem{kendall79}
S.~M. Kendall and A.~Stuart, \emph{The Advanced Theory of Statistics},
  2nd~ed.\hskip 1em plus 0.5em minus 0.4em\relax New York, NY, USA: Macmillan,
  1979.

\bibitem{conte00}
E.~Conte, A.~D. Maio, and C.~Galdi, ``Signal detection in compound-gaussian
  noise: {N}eyman-{P}earson and {CFAR} detectors,'' \emph{IEEE Trans. Signal
  Process.}, vol.~48, no.~2, pp. 419--428, Feb. 2000.

\bibitem{kay98}
S.~M. Kay, \emph{Fundamentals of Statistical Signal Processing: Detection
  Theory}, 2nd~ed.\hskip 1em plus 0.5em minus 0.4em\relax Upper Saddle River,
  NJ, USA: Prentice Hall PTR, 1998.

\bibitem{Almeida19}
F.~D.~A. {García}, H.~R.~C. Mora, and N.~V.~O. {Garzón}, ``{GLRT} detection
  of nonfluctuating targets in background noise using phased arrays,'' in
  \emph{Proc. 15th IEEE International Conference on Wireless and Mobile
  Computing, Networking and Communications ({WIMOB})}, Barcelona, Spain, Oct.
  2019, pp. 1--8.

\bibitem{Haykin92}
S.~S. Haykin and A.~O. Steinhardt, \emph{Adaptive Radar Detection and
  Estimation}, 1st~ed.\hskip 1em plus 0.5em minus 0.4em\relax New Jersey, NJ,
  USA: J. Wiley, 1992.

\bibitem{kelly86}
E.~J. Kelly, ``An adaptive detection algorithm,'' \emph{IEEE Trans. Aerosp.
  Electron. Syst.}, vol. AES-22, no.~2, pp. 115--127, Mar. 1986.

\bibitem{Reed74}
I.~S. {Reed}, J.~D. {Mallett}, and L.~E. {Brennan}, ``Rapid convergence rate in
  adaptive arrays,'' \emph{IEEE Trans. Aerosp. Electron. Syst.}, vol. AES-10,
  no.~6, pp. 853--863, Nov. 1974.

\bibitem{bose96}
S.~Bose and A.~O. Steinhardt, ``Optimum array detector for a weak signal in
  unknown noise,'' \emph{IEEE Trans. Aerosp. Electron. Syst.}, vol.~32, no.~3,
  pp. 911--922, Jul. 1996.

\bibitem{besson17}
O.~Besson, A.~Coluccia, E.~Chaumette, G.~Ricci, and F.~Vincent, ``Generalized
  likelihood ratio test for detection of gaussian rank-one signals in gaussian
  noise with unknown statistics,'' \emph{IEEE Trans. Signal Process.}, vol.~65,
  no.~4, pp. 1082--1092, Feb. 2017.

\bibitem{Pulsone01}
N.~B. {Pulsone} and C.~M. {Rader}, ``Adaptive beamformer orthogonal rejection
  test,'' \emph{IEEE Trans. Signal Process.}, vol.~49, no.~3, pp. 521--529,
  Mar. 2001.

\bibitem{Robey92}
F.~C. {Robey}, D.~R. {Fuhrmann}, E.~J. {Kelly}, and R.~{Nitzberg}, ``A {CFAR}
  adaptive matched filter detector,'' \emph{IEEE Trans. Aerosp. Electron.
  Syst.}, vol.~28, no.~1, pp. 208--216, Jan. 1992.

\bibitem{Zhang18}
S.~{Zhang}, C.~{Guo}, T.~{Wang}, and W.~{Zhang}, ``{ON–OFF} analog
  beamforming for massive {MIMO},'' \emph{IEEE Trans. Veh. Technol.}, vol.~67,
  no.~5, pp. 4113--4123, Jan. 2018.

\bibitem{Huber12}
S.~{Huber}, M.~{Younis}, A.~{Patyuchenko}, G.~{Krieger}, and A.~{Moreira},
  ``Spaceborne reflector {SAR} systems with digital beamforming,'' \emph{IEEE
  Trans. Aerosp. Electron. Syst.}, vol.~48, no.~4, pp. 3473--3493, Oct. 2012.

\bibitem{Axelsson03}
S.~R.~J. {Axelsson}, ``Noise radar for range/doppler processing and digital
  beamforming using low-bit {ADC},'' \emph{IEEE Trans. Geosci. Remote Sens.},
  vol.~41, no.~12, pp. 2703--2720, Dec. 2003.

\bibitem{Zhu17}
D.~{Zhu}, B.~{Li}, and P.~{Liang}, ``A novel hybrid beamforming algorithm with
  unified analog beamforming by subspace construction based on partial {CSI}
  for massive {MIMO-OFDM} systems,'' \emph{IEEE Trans. Commun.}, vol.~65,
  no.~2, pp. 594--607, Nov. 2017.

\bibitem{richards10}
M.~A. Richards, J.~Scheer, W.~A. Holm, and W.~L. Melvin, \emph{Principles of
  Modern Radar: Basic Principles}, 1st~ed.\hskip 1em plus 0.5em minus
  0.4em\relax West Perth, WA, Australia: SciTech, 2010.

\bibitem{Weinberg17trans}
G.~V. {Weinberg}, ``Noncoherent radar detection in correlated {P}areto
  distributed clutter,'' \emph{IEEE Trans. Aerosp. Electron. Syst.}, vol.~53,
  no.~5, pp. 2628--2636, Oct. 2017.

\bibitem{Weinberg19}
G.~V. {Weinberg} and C.~{Tran}, ``Noncoherent detector threshold determination
  in correlated {P}areto distributed clutter,'' \emph{IEEE Geosci. Remote Sens.
  Lett.}, vol.~16, no.~3, pp. 372--376, Mar. 2019.

\bibitem{Weinberg17letter}
G.~V. {Weinberg}, ``Minimum-based sliding window detectors in correlated
  {P}areto distributed clutter,'' \emph{IEEE Geosci. Remote Sens. Lett.},
  vol.~14, no.~11, pp. 1958--1962, Nov. 2017.

\bibitem{skolnik01}
M.~I. Skolnik, \emph{Introduction to Radar Systems}, 3rd~ed.\hskip 1em plus
  0.5em minus 0.4em\relax Ney York, NY, USA: McGraw-Hill, 2001.

\bibitem{papoulis02}
A.~Papoulis, \emph{Probability, Random Variables, and Stochastic Processes},
  4th~ed.\hskip 1em plus 0.5em minus 0.4em\relax Ney York, NY, USA:
  McGraw-Hill, 2002.

\bibitem{patnaik49}
P.~B. Patnaik, ``The non-central $\chi^2$ and \textit{F}-distributions and
  their applications,'' \emph{Biometrika}, vol.~36, no.~1, pp. 202--232, Jun.
  1949.

\bibitem{phillips82}
P.~C.~B. Phillips, ``The true characteristic function of the {F}
  distribution,'' \emph{Biometrika}, vol.~69, no.~1, p. 261–264, Apr. 1982.

\bibitem{Olver10}
F.~W.~J. Olver, D.~W. Lozier, R.~F. Boisvert, and C.~W. Clark, \emph{NIST
  Handbook of Mathematical Functions}, 1st~ed.\hskip 1em plus 0.5em minus
  0.4em\relax Washington, DC: US Dept. of Commerce: National Institute of
  Standards and Technology (NIST), 2010.

\bibitem{Bulgren71}
W.~G. Bulgren, ``On representations of the doubly non-central {F}
  distribution,'' \emph{J. Amer. Statist.}, vol.~66, no. 333, pp. 184--186,
  Mar. 1971.

\bibitem{Garcia19}
F.~D.~A. García, A.~C.~F. Rodriguez, G.~Fraidenraich, and J.~C.~S. {Santos
  Filho}, ``{CA-CFAR} detection performance in homogeneous {W}eibull clutter,''
  \emph{IEEE Geosci. Remote Sens. Lett.}, vol.~16, no.~6, pp. 887--891, Jun.
  2019.

\bibitem{Rahama18}
Y.~{Abo Rahama}, M.~H. {Ismail}, and M.~S. {Hassan}, ``On the sum of
  independent {Fox's $H$} -function variates with applications,'' \emph{IEEE
  Trans. Veh. Technol.}, vol.~67, no.~8, pp. 6752--6760, Aug. 2018.

\bibitem{carlos17}
C.~R.~N. da~Silva, E.~J. Leonardo, and M.~D. Yacoub, ``Product of two envelopes
  taken from $\alpha-\mu$, $\kappa-\mu$ and $\eta-\mu$ distributions,''
  \emph{IEEE Trans. Commun.}, vol.~PP, no.~99, pp. 1--1, Mar. 2017.

\bibitem{hfox}
C.~H.~M. {de Lima}, H.~{Alves}, and P.~H.~J. {Nardelli}, ``Fox {$H$-function}:
  A study case on variate modeling of dual-hop relay over {W}eibull fading
  channels,'' in \emph{2018 IEEE Wireless Communications and Networking
  Conference (WCNC)}, Apr. 2018, pp. 1--5.

\bibitem{AlmeidaIET20}
F.~D.~A. {García}, H.~R.~C. {Mora}, G.~{Fraidenraich}, and J.~C.~S. {Santos
  Filho}, ``Alternative representations for the probability of detection of
  non-fluctuating targets,'' \emph{Electron. Lett.}, vol.~56, no.~21, pp.
  1136--1139, Oct. 2020.

\bibitem{hay95}
N.~T. Hai and H.~M. Srivastava, ``The convergence problem of certain multiple
  {M}ellin-{B}arnes contour integrals representing {H}-functions in several
  variables,'' \emph{Computers {\&} Mathematics with Applications}, vol.~29,
  no.~6, pp. 17--25, 1995.

\bibitem{abramowitz72}
M.~Abramowitz and I.~A. Stegun, \emph{Handbook of Mathematical Functions with
  Formulas, Graphs, and Mathematical Tables}, 10th~ed.\hskip 1em plus 0.5em
  minus 0.4em\relax Washington, DC: US Dept. of Commerce: National Bureau of
  Standards, 1972.

\bibitem{prudnikov03}
A.~P. Prudnikov, Y.~A. Bry{\v c}kov, and O.~I. Mari{\v c}ev, \emph{Integral and
  Series: {V}ol. 3}, 2nd~ed., Fizmatlit, Ed.\hskip 1em plus 0.5em minus
  0.4em\relax Moscow, Russia: Fizmatlit, 2003.

\bibitem{fubibi07}
G.~{Fubini}, ``\BIBforeignlanguage{Italian}{{Sugli integrali multipli.}}''
  \emph{\BIBforeignlanguage{Italian}{{Rom. Acc. L. Rend. (5)}}}, vol.~16,
  no.~1, pp. 608--614, 1907.

\bibitem{Mathematica}
\BIBentryALTinterwordspacing
{Wolfram Research, Inc. (2018)}, \emph{Wolfram Research}, Accessed: Sept. 19,
  2020. [Online]. Available: \url{http://functions.wolfram.com}
\BIBentrySTDinterwordspacing

\bibitem{alhennawi16}
H.~R. Alhennawi, {M. M. H. E. Ayadi}, M.~H. Ismail, and H.~A.~M. Mourad,
  ``Closed-form exact and asymptotic expressions for the symbol error rate and
  capacity of the {H}-function fading channel,'' \emph{IEEE Trans. Veh.
  Technol.}, vol.~65, no.~4, pp. 1957--1974, Apr. 2016.

\bibitem{yilmaz09}
F.~Yilmaz and M.~S. Alouini, ``Product of the powers of generalized
  nakagami-$m$ variates and performance of cascaded fading channels,'' in
  \emph{Proc. IEEE Global Telecommun. Conf. (GLOBECOM)}, Abu Dhabi, UAE, Nov.
  2009, pp. 1--8.

\bibitem{AlmeidaIEEE20}
F.~D.~A. {García}, H.~R.~C. {Mora}, G.~{Fraidenraich}, and J.~C.~S. {Santos
  Filho}, ``Square-law detection of exponential targets in
  {W}eibull-distributed ground clutter,'' \emph{IEEE Geosci. Remote Sens.
  Lett.}, to be published, doi:
  \href{https://doi.org/10.1109/LGRS.2020.3009304}{\textcolor{black}{10.1109/LGRS.2020.3009304}}.

\bibitem{Kreyszig10}
E.~Kreyszig, \emph{Advanced Engineering Mathematics}, 10th~ed.\hskip 1em plus
  0.5em minus 0.4em\relax New Jersey, NJ, USA: John Wiley \& Sons, 2010.

\bibitem{cochran34}
W.~G. Cochran, ``The distribution of quadratic forms in a normal system, with
  applications to the analysis of covariance,'' \emph{Proc. Camb. Phil. Soc.},
  vol.~30, no.~2, p. 178–191, 1934.

\bibitem{springer79}
M.~D. Springer, \emph{The Algebra of Random Variables}.\hskip 1em plus 0.5em
  minus 0.4em\relax New York, NY, USA: Wiley, 1979.

\end{thebibliography}

\end{document}